\documentclass[pre,notitlepage,superscriptaddress,showpacs,showkeys,longbibliography,twocolumn]{revtex4-1}
\usepackage{latexsym,amssymb,amsfonts,mathrsfs,amsmath,bm}
\usepackage{graphicx}
\usepackage[usenames,dvipsnames]{xcolor}
\usepackage{soul}
\usepackage{float}
\usepackage{bbold}
\usepackage[linktocpage,colorlinks=true,linkcolor=blue,citecolor=blue,breaklinks=true,urlcolor=blue]{hyperref}

\newcommand{\bu}{{\bf u}}

\newcommand{\bx}{{\bf x}}

\newcommand{\RE}[1]{\mathbb{R}{e} [#1]}


\begin{document}

\title{Flexible filaments buckle into helicoidal shapes in strong compressional flows}  

\author{Brato Chakrabarti}
\affiliation{Department of Mechanical and Aerospace Engineering, University of California San Diego, 9500 Gilman Drive, La Jolla, CA 92093, USA}
\author{Yanan Liu}
\email[E-mail address: ]{yanan.liu@nwu.edu.cn}
\affiliation{Laboratoire de Physique et M\'ecanique des Milieux H\'et\`erog\`enes, UMR 7636, ESPCI Paris, PSL Research University, CNRS, Universit\'e Paris Diderot, Sorbonne Universit\'e, 10 rue Vauquelin, 75005 Paris, France}
\author{John LaGrone}
\author{Ricardo Cortez}
\author{Lisa Fauci}
\affiliation{Department of Mathematics, Tulane University, 6823 St.\ Charles Avenue, New Orleans, LA 70118, USA}
\author{Olivia du Roure}
\affiliation{Laboratoire de Physique et M\'ecanique des Milieux H\'et\`erog\`enes, UMR 7636, ESPCI Paris, PSL Research University, CNRS, Universit\'e Paris Diderot, Sorbonne Universit\'e, 10 rue Vauquelin, 75005 Paris, France}
\author{David Saintillan}
\email[E-mail address: ]{dstn@ucsd.edu}
\affiliation{Department of Mechanical and Aerospace Engineering, University of California San Diego, 9500 Gilman Drive, La Jolla, CA 92093, USA}
\author{Anke Lindner}
\affiliation{Laboratoire de Physique et M\'ecanique des Milieux H\'et\`erog\`enes, UMR 7636, ESPCI Paris, PSL Research University, CNRS, Universit\'e Paris Diderot, Sorbonne Universit\'e, 10 rue Vauquelin, 75005 Paris, France}


\begin{abstract}
\vspace{0.1cm}
\noindent \textbf{\textsf{The occurrence of coiled or helical morphologies is common in nature, from plant roots to DNA packaging into viral capsids, as well as in applications such as oil drilling processes.\ In many examples, chiral structures result from the buckling of a straight fiber either with intrinsic twist or to which end moments have been applied in addition to compression forces.\ Here, we elucidate a generic way to form regular helicoidal shapes from achiral straight filaments transported in viscous flows with free ends.\ Through a combination of experiments using fluorescently labeled actin filaments in microfluidic divergent flows and of two distinct sets of numerical simulations, we demonstrate the robustness of helix formation.\ A nonlinear stability analysis is performed and explains the emergence of such chiral structures from the nonlinear interaction of perpendicular planar buckling modes, an effect that solely requires a strong compressional flow, independent of the exact nature of the fiber or type of flow field.\ The fundamental mechanism for the uncovered morphological transition  and characterization of the emerging conformations advance our understanding of several biological and industrial processes and can also be exploited for the controlled microfabrication of chiral objects.\\ $\,$}}
\end{abstract}

\maketitle

\noindent The interaction of flexible structures with viscous flows is central for a wide range of important biological and industrial processes, such as the swimming of flagellated microorganisms \cite{lauga2016bacterial}, the squeezing of red blood cells though microcapillaries \cite{abkarian2016}, mitotic spindle formation and positioning during cell division \cite{Nazockdast2017}, or the design of soft micromechanical sensors \cite{chen2014stimuli}. In all cases, the dynamics and morphologies of the flexible objects subject to viscous stresses underly the observed or desired properties.

Particularly rich dynamics occur when flexible objects are elongated \cite{duroure2019fibers}, where complex morphologies emerge as nonuniform viscous stresses overcome structural rigidity.\ Elongated shapes not only lead to drag anisotropy but also result in high deformability as bending rigidity is strongly dependent on geometry in addition to material properties.\ Buckling instabilities of freely transported flexible fibers have been predicted, simulated and experimentally observed in shear as well as stagnation point flows \cite{becker2001instability, young2007stretch, kantsler2012fluctuations, gugli2012buckling,lindner2015elastic, duroure2019fibers} and result in characteristic two-dimensional buckling modes.\ Under sedimentation, flexible fibers spontaneously adopt a steady $U$ shape independent of initial conditions \cite{li2013sedimentation}.  Flexible filaments can also be compressed into more compact, three-dimensional conformations under stronger forcing, and fiber coiling from the ends has been reported in shear \cite{forgacs1959particle, harasim2013direct, liu2018morphological, nguyen2014hydrodynamics, lagrone2019complex}.\ However, these morphologies are typically irregular and random, except for very specific initial conditions where, for instance, knot formation has been observed \cite{Kuei_2015}.

\begin{figure*}[t]
	\centering 
	\includegraphics[width=0.955\linewidth]{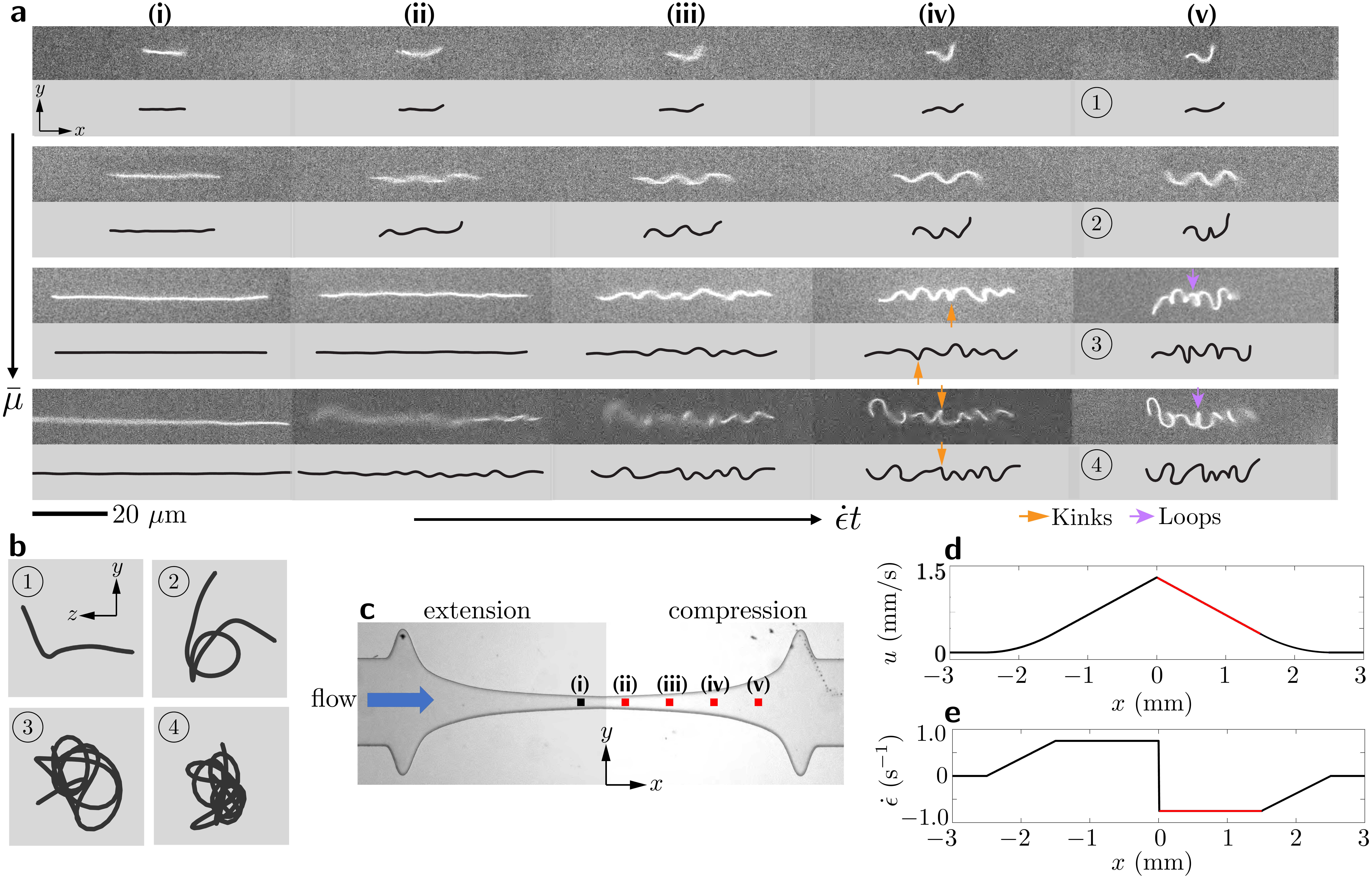}\vspace{-0.3cm}
	\caption{\textbf{\textsf{Buckling conformations of Brownian filaments in compressional flow.}}  \textbf{\textsf{a,}} Snapshots of evolving filament morphologies from experiments and Brownian simulations compared at the same dimensionless time $\dot{\epsilon}t$, measured from the instant the filament enters the compressional region.\ Vertical panels correspond to increasing values of the elastoviscous number in the range $\bar{\mu}\sim  2 \times 10^3-10^6$.\ The filament is initially extended and aligned with the centerline (i), and it starts buckling after experiencing compression. For sufficiently large $\bar{\mu}$, 3D shapes emerge, as evidenced by kinks (orange arrows) and loops (purple arrows) in the 2D projections.\ Also see Supplementary Videos 1--5. \textbf{\textsf{b,}} Projections of simulated filament conformations corresponding to (v) in the cross-sectional plane of the channel, highlighting the 3D helicoidal nature of the morphologies at large $\bar{\mu}$. \textbf{\textsf{c,}} Geometry of the optimized hyperbolic microfluidic channel used in experiments, with markers indicating the positions where the snapshots shown in \textbf{\textsf{a}} were taken. \textbf{\textsf{d-e,}} Axial velocity $u$ and strain rate $\dot{\epsilon}$ as functions of streamwise position $x$ along the channel centerline where filaments are transported.\ A constant strain rate $\dot{\epsilon}$ occurs over a given distance, and measurements are made in the compressional region highlighted in red. \vspace{-0.35cm}}
	\label{fig:snapshots}
\end{figure*}

Here, we report on the surprising finding that a freely suspended straight flexible filament can buckle into a helical shape in a purely compressional flow.\ The formation of such regular three-dimensional conformations  \cite{silverberg20123d, svenvsek2008confined, klug2005three, miller2015buckling} typically requires applying end moments in addition to compressive forces \cite{goriely2000nonlinear,van2000helical,antman2005problems}.\ The phenomenon discussed here thus stands out from  classical helical buckling in that the filament spontaneously adopts a chiral helicoidal morphology in the absence of any intrinsic twist or external moments. 

 We elucidate this generic morphological transition through a combination of experiments, simulations and theoretical modeling.\ To induce and visualize buckling in experiments, we flowed fluorescently labeled actin filaments through a convergent-divergent hyperbolic microfluidic channel  
specially designed and optimized to provide uniform extension and compression rates over large distances while ensuring a long residence time for the filaments \cite{zografos2016microfluidic,liu2019optimized}.\ These experiments are complemented by two sets of very different simulations.\ In the first model closely mimicking the experimental conditions, we performed Langevin simulations of inextensible Euler-Bernoulli beams placed in a two-dimensional flow field and subject to thermal fluctuations \cite{liu2018morphological}.\ In the second model, we simulated non-Brownian elastic fibers composed of surface nodes connected by a network of springs providing structural rigidity and bending resistance \cite{nguyen2014hydrodynamics, lcf2019} in an axisymmetric channel \cite{lagrone2019complex}.\ 
As shown below, both types of simulations, as well as seminal simulations by Chelakkot et al.~\cite{chelakkot2012flow}, recapitulate the helix formation seen in experiments, pointing to a very generic transition that only requires a strong compressional flow as we rationalize below using a nonlinear stability analysis.\ Our findings highlight a new mechanism by which a one-dimensional object can buckle into a chiral helicoidal shape under viscous loading.\ This mechanism remained undiscovered until now as typical experimental setups in stagnation point flows do not allow for sufficiently strong compression rates or long residence times, and as past theoretical analyses have been limited to two dimensions.\ Our results also underscore the robustness of this phenomenon, which occurs independent of the presence of thermal fluctuations and across very different flow environments.



\vspace{0.2cm}
\noindent \textbf{\textsf{Strong compressional flows induce helical buckling}}\\
  Typical buckling events in experiments and Brownian simulations are shown in Fig.~\ref{fig:snapshots}\textbf{\textsf{a}} for increasing values of the elastoviscous number $\bar{\mu}$ \cite{duroure2019fibers}, a dimensionless measure of compression rate whose definition we specify below.\ We focus here on the dynamics in the divergent part of a microfluidic hyperbolic channel (see Fig.~\ref{fig:snapshots}\textbf{\textsf{c-e}}) where compression at constant strain rate $\dot{\epsilon}$ occurs, with the convergent part mainly serving to align and prestretch the filaments before measurements begin.\ As the filaments enter the compressional region (column (i)), they are indeed mostly straight as thermal shape fluctuations have been largely suppressed \cite{kantsler2012fluctuations}.\ Snapshots at increasing values of the dimensionless time $\dot{\epsilon}t$ in panels (ii) to (v) show the growth of deformations with distinct emergent morphologies for increasing elastoviscous numbers (top to bottom).\ In relatively weak flows (first row), deformations are mostly planar and resemble those seen in past studies in stagnation point flows \cite{kantsler2012fluctuations,manikantan2015buckling, quennouz2015transport}. As $\bar{\mu}$ is increased in subsequent rows, more complex shapes emerge that are fully three-dimensional, as evidenced by the blurriness of some parts of the filaments in the experimental images due to deformations out of the focal plane.\ Another indicator of three-dimensionality is the presence of apparent kinks (orange arrows) in the 2D images, which must result from the projection of 3D shapes. In some cases, actual loops (purple arrows) can be observed and strongly hint at helicoidal shapes.\ This is confirmed in Fig.~\ref{fig:snapshots}\textbf{\textsf{b}}, showing simulated \textcolor {black} {Brownian} filament projections in the cross-sectional plane, where these loops are now clearly visible. The number of loops along the filament increases with $\bar{\mu}$ as higher unstable buckling modes become excited. The emerging coiled structures have no preferred chirality as expected from symmetry, and in some cases reversals in the handedness occur at topological perversions along the contour length. As the filament is transported downstream, the helix is further compressed by the flow until it exits the compressional region and is finally allowed to relax.

\begin{figure}[t]
	\centering \vspace{0.2cm}
	\includegraphics[width=0.48\textwidth]{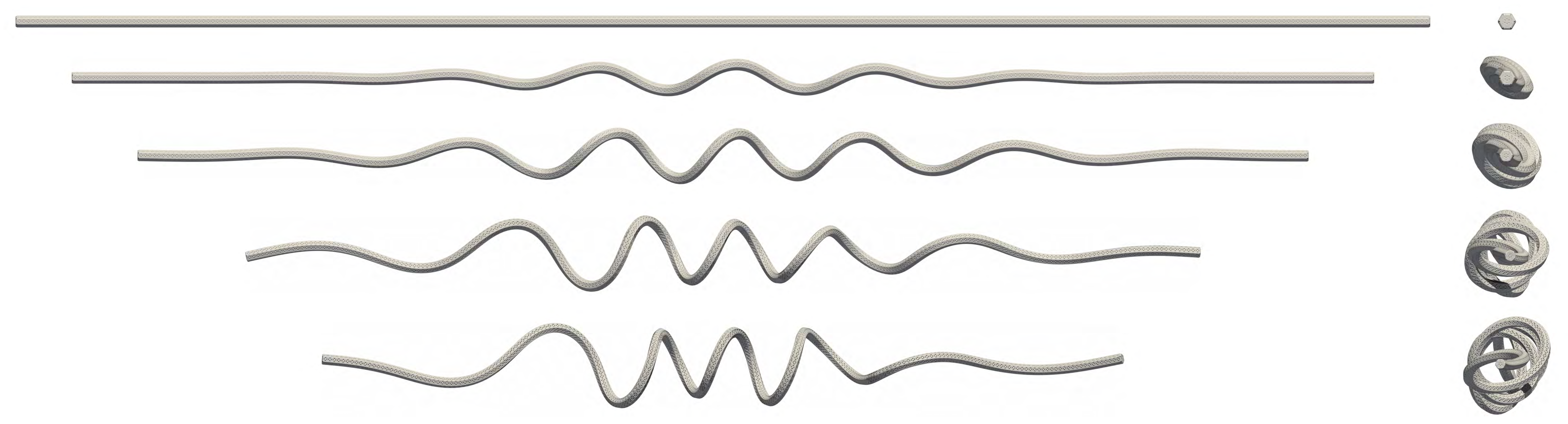}\vspace{-0.15cm}
	\caption{\textbf{\textsf{Helical buckling of a non-Brownian filament.}} Typical buckling sequence in a simulation of a non-Brownian filament with $\bar{\mu}=6.5 \times 10^4$.  In the simulation shown, deformations first occur in a 2D plane before the 3D helical shape develops.\ Deformations also tend to be largest near the center of the filament.\ Also see Supplementary Video 8. \vspace{-0.3cm}}
	\label{fig:SimB}
\end{figure}

\begin{figure*}
	\centering
	\includegraphics[width=0.98\linewidth]{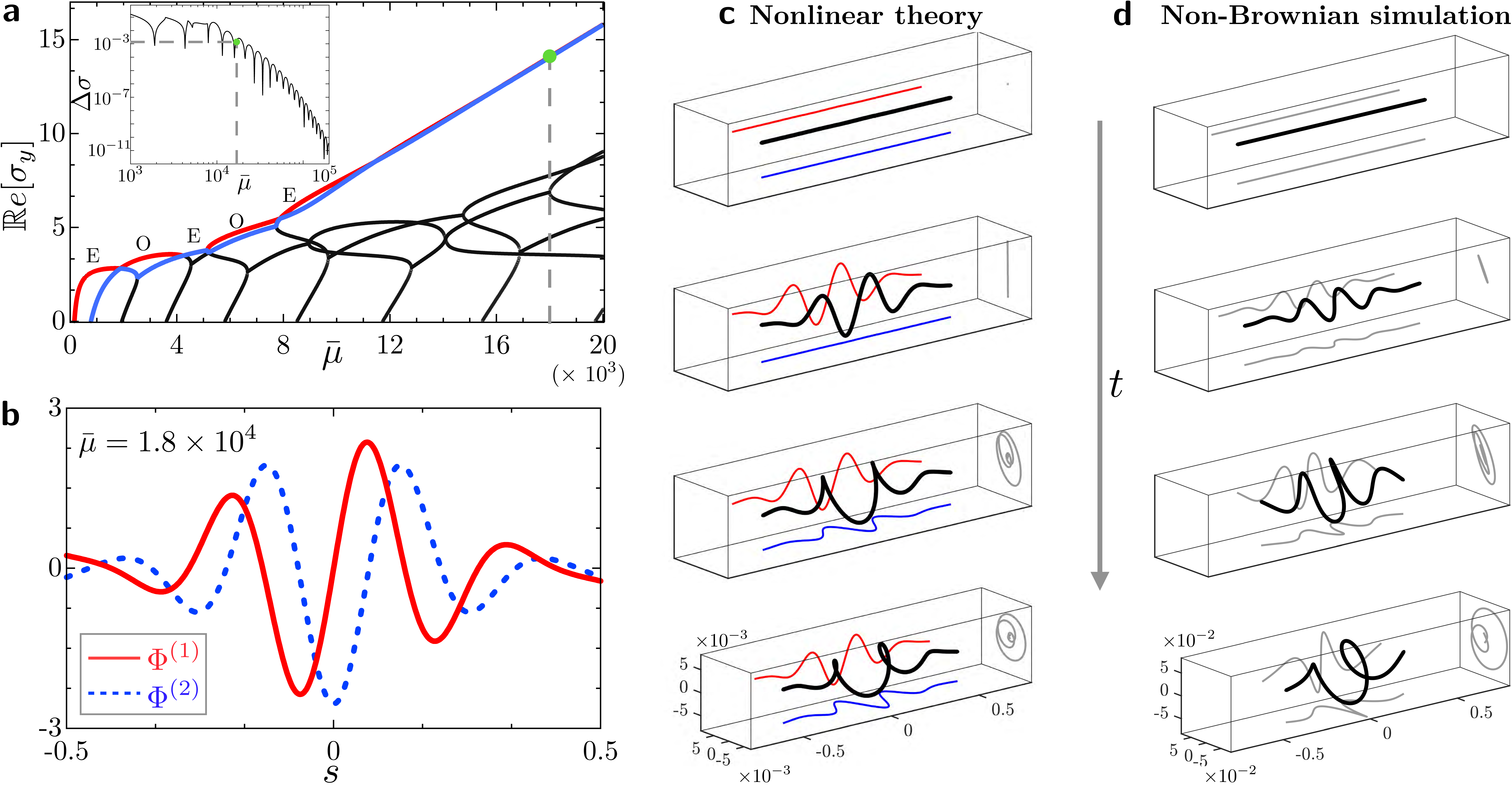}\vspace{-0.2cm}
	\caption{\textbf{\textsf{Stability analysis and rationale for helical shapes.}} \textsf{\textbf{a,}} Growth rates of unstable planar eigenmodes plotted vs.\ elastoviscous number. As $\bar{\mu}$ increases, additional modes become unstable. The two dominant eigenvalues are colored in red and blue and labeled as either even (E) or odd (O) functions of $s$. Inset: difference $\Delta\sigma$ between the two most unstable eigenvalues as a function of $\bar{\mu}$, showing convergence in strong flows. \textbf{\textsf{b,}} Eigenmodes associated with the two largest eigenvalues for a value of $\bar{\mu}=1.8\times 10^{4}$ corresponding to the green dot in \textsf{\textbf{a}}. The dominant mode $\Phi^{(1)}$ is odd, whereas $\Phi^{(2)}$ is even. \textbf{\textsf{c,}} Snapshots from a time sequence predicted by the nonlinear model, showing the formation of a 3D coiled conformation from the superposition of adjacent even and odd planar eigenmodes in orthogonal planes.\ See Supplementary Video~9 for an animation. \textbf{\textsf{d,}} Buckling sequence in a non-Brownian simulation at the same value of $\bar{\mu}$, also showing the growth of a helix from an initially planar buckling mode.\ Also see Supplementary Video 7. }\vspace{-0.3cm}
	\label{fig:Eigmode}
\end{figure*}

\textcolor {black} {Simulations of a non-Brownian fiber} in Fig.~\ref{fig:SimB} are consistent with these observations and provide a cleaner picture of the buckling process.\ In absence of thermal fluctuations, deformations are typically concentrated near the center of the filament, with the filament ends remaining mostly straight and aligned with the flow axis.\ In simulations at moderate flow strengths, we find that deformations first occur in a two-dimensional plane before three-dimensional effects kick in and lead to the helix formation.\ This curious sequence of events, which we elucidate below, disappears in very strong flows where three-dimensional shapes emerge almost instantly.\ As in the Brownian case, shape perversions occasionally arise along the filament and cause handedness reversals \cite{goriely1998spontaneous,goriely2000nonlinear}.

\vspace{0.35cm}
\noindent \textbf{\textsf{Helical shapes stem from interacting planar modes} }\\
 We proceed to explain the emergence of helical morphologies using a 3D weakly nonlinear stability analysis.\ Previous 2D linear analyses in planar flows have been very successful at predicting the onset of buckling and subsequent mode shapes \cite{becker2001instability,young2007stretch,manikantan2015buckling,lindner2015elastic}.\ Here, we show how the interaction of pairs of planar eigenmodes growing in different planes is responsible for the observed 3D helices.\ 

 In absence of thermal fluctuations, filament dynamics are governed by the interplay of viscous forces exerted by the flow and internal elasticity.\ This balance is quantified by the dimensionless elastoviscous number $\bar{\mu}$, comparing the characteristic timescale for elastic relaxation of a bending mode to the timescale of the imposed flow \cite{liu2018morphological}.\ It is defined as $\bar{\mu} = 8 \pi \mu \dot{\epsilon}L^4/Bc$ in terms of the solvent viscosity $\mu$, applied compression rate $\dot{\epsilon}$, filament contour length $L$ and bending rigidity $B$ and shows a strong dependence on length.\ The constant $c=-\ln(\alpha^2e)$ is a dimensionless slenderness parameter where $\alpha$ is the filament aspect ratio.\ When the filament is small enough to experience Brownian motion, the balance of fluctuations and bending rigidity results in a persistence length $\ell_p=B/k_\mathrm{B}T$ where $k_\mathrm{B}T$ is the thermal energy.\ In this case, the ratio  $\ell_p/L$ quantifies the amplitude of fluctuations, with the limit of $\ell_p/L  \to \infty$ describing Brownian rigid fibers.\ In past studies of viscous buckling, the role of thermal fluctuations was shown to be limited to triggering the instability while smoothing the buckling transition \cite{kantsler2012fluctuations,manikantan2015buckling,liu2018morphological}.\ We neglect fluctuations in our theory, though their effect will be addressed in simulations.

In the reference frame of the translating filament, the local flow field is well approximated by a planar compressional flow in experiments and Brownian simulations, and by a uniaxial compressional flow in non-Brownian simulations; we focus here on the former case and thus take the dimensionless flow field to be $\bu_\infty = (-x,y,0)$.\ In the base state, the filament is straight with its center at the stagnation point and its axis aligned with the direction of compression: $\mathbf{x}_0(s)=s\hat{\mathbf{x}}$ with $s\in[-0.5,0.5]$. Its motion is described using local slender-body theory for low-Reynolds-number hydrodynamics  \cite{batchelor1970},
\begin{equation}\label{eq:lsbt}
\bar{\mu}\,(\dot{\bx} - \bu_\infty) = \left(\mathbb{1} + \bx_s \bx_s \right) \cdot \left[(T \bx_s)_s - \bx_{ssss}\right], 
\end{equation}
where the elastoviscous number $\bar{\mu}$ appears as the sole control parameter.\ Indices in Eq.~(\ref{eq:lsbt}) denote differentiation with respect to arclength, with $\mathbf{x}_s$ describing the local tangent vector.\ The scalar $T(s)$ is the internal tension that enforces filament inextensibility. Eq.~(\ref{eq:lsbt}) is accompanied by force- and moment-free boundary conditions: $\bx_{sss}=\bx_{ss}=T=0$ at $s=\pm1/2$.\ In the base state, the compressional flow induces a parabolic tension profile $T_0(s) = \frac{1}{4}\bar{\mu}\left(s^2 - \frac{1}{4}\right)$ typical of undeformed filaments in linear flows \cite{becker2001instability,young2007stretch}.\ 

The straight configuration is perturbed as
 $\mathbf{x}(s,t)=(s,h_y,h_z)$, where $h_y(s,t)$ and $h_z(s,t)$ are in-plane $(x,y)$ and out-of-plane $(x,z)$ shape perturbations, respectively, and are assumed to be small $\mathcal{O}(\varepsilon)$ quantities.\  We first perform a linear analysis and simplify Eq.~(\ref{eq:lsbt}) as
\begin{equation}\label{eq:linsb}
\bar{\mu}\,(\dot{\bx} - \bu_\infty) =T_{0} \bx_{ss} + 2 T_{0,s} \bx_{s} - \bx_{ssss} + \mathcal{O}(\varepsilon^2),
\end{equation}
where the velocity field experienced by the filament is $\mathbf{u}_\infty=(-s,h_y,0)$.
We seek normal-mode perturbations of the form $\{h_y,h_z\} = \{\Phi_y(s),\Phi_z(s)\}\exp(\sigma t)$, where $\Phi_y$ and $\Phi_z$ are in- and out-of-plane mode shapes and $\sigma$ is the complex growth rate.\ Inserting this ansatz into Eq.~(\ref{eq:linsb}) yields two eigenvalue problems in the $y$ and $z$ directions:
\begin{align}
\bar{\mu}\, (\sigma-1)\, \Phi_y &= \mathcal{L}[\Phi_y], \label{eq:eigx} \\
\bar{\mu}\, \sigma\, \Phi_z &= \mathcal{L}[\Phi_z] \label{eq:eigy} ,
\end{align}
where $\mathcal{L}$ is the differential operator $\mathcal{L}[\Phi]=T_0(s)\Phi_{ss}+T_{0,s}(s)\,\Phi_s-\Phi_{ssss}$. Inspection of Eqs.~(\ref{eq:eigx})--(\ref{eq:eigy}) shows that the eigenvalue problems in the two orthogonal planes are uncoupled and thus have their own growth rates $(\sigma_y,\sigma_z)$.\ Incidentally, the two eigenvalue problems are found to be identical under the transformation $\sigma_z=\sigma_y-1$.\ This points to a key aspect of the eigenspectrum: for a given value of $\bar{\mu}$, in- and out-of-plane mode shapes are identical but have offset growth rates.\ Out-of-plane deformations grow slightly more slowly as a consequence of the 2D nature of the flow; both growth rates would be identical in uniaxial flow. 

The eigenvalue problem of Eq.~(\ref{eq:eigx}) was solved numerically using a Chebyshev spectral collocation method with boundary conditions $\Phi_{yss} = \Phi_{ysss} = 0$ at $s=\pm1/2$, and pertinent results are summarized in Fig.~\ref{fig:Eigmode}.\ The growth rates $\RE{\sigma_y}$ of unstable modes are plotted versus $\bar{\mu}$ in Fig.~\ref{fig:Eigmode}\textbf{\textsf{a}}.\ In very weak flows, all modes are stable with negative growth rates.\ As the elastoviscous number is increased, a supercritical pitchfork bifurcation occurs giving rise to the first onset of buckling.\ In agreement with past planar analyses \cite{young2007stretch,manikantan2015buckling,quennouz2015transport}, the first buckling threshold is found to be $\smash{\bar{\mu}_{c}\approx 153.2}$  with an even mode shape ($\Phi(-s)=\Phi(s)$) resembling the canonical $C$ shape typical of Euler buckling. At yet larger values of $\bar{\mu}$, higher-order buckling modes with larger wavenumbers are excited and can become unstable, leading to the complex eigenspectrum of Fig.~\ref{fig:Eigmode}\textbf{\textsf{a}}. Three essential features stand out: (i) at large elastoviscous numbers, the first two eigenvalues $\smash{\{\sigma_y^{(1)},\sigma_y^{(2)}\}}$ dominate the spectrum and the corresponding eigenmodes are expected to dictate the emergent morphologies; (ii) these two dominant  eigenmodes always come in an odd--even pair, i.e., if $\Phi^{(1)}$ is odd then $\Phi^{(2)}$ is even and vice-versa; (iii) the difference in growth rate between these two dominant modes becomes negligible in strong flows.\ This last point is made clear in the inset of Fig.~\ref{fig:Eigmode}\textbf{\textsf{a}}, where the difference $\Delta \sigma$ between the two growth rates is seen to decay rapidly with $\bar{\mu}$. 

\begin{figure*}
	\centering 
	\includegraphics[width=0.96\linewidth]{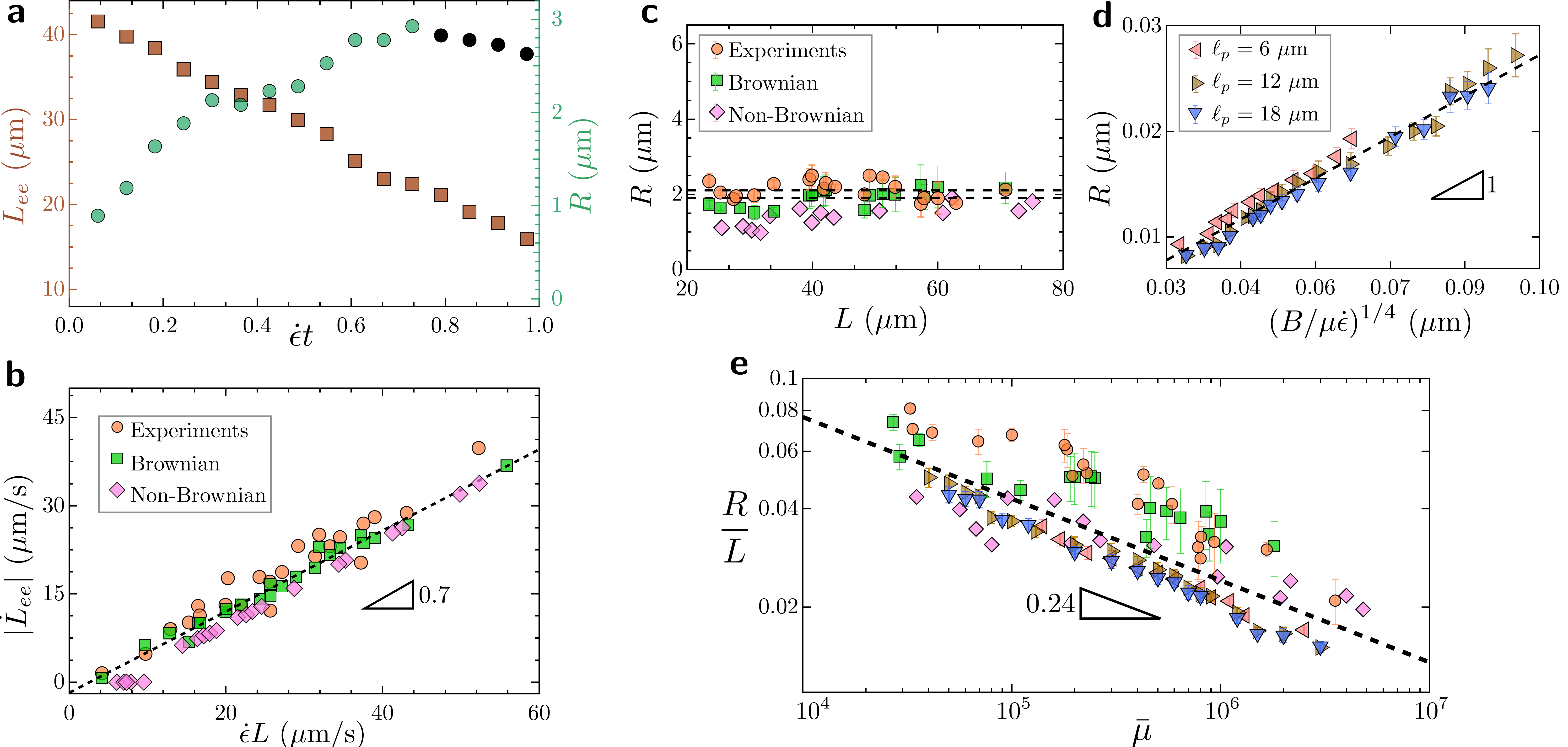}\vspace{-0.15cm}
	\caption{\textbf{\textsf{Temporal dynamics and final helix radius. a,}}  Evolution of the end-to-end distance $L_{ee}$ and  coiling radius $R$ as functions of dimensionless time $\dot{\epsilon}t$ in an experiment with $L=42.1\,\mu$m and $\dot{\epsilon}=0.61\,$s$^{-1}$ (see Supplementary Fig.~4 for corresponding data from simulations). The range used to estimate the final helix radius is shown in black. \textbf{\textsf{b,}} Compression speed of the helix estimated as the rate of change of the end-to-end distance, showing a linear dependence on $\dot{\epsilon}L$ with a slope of $\sim0.7$. \textbf{\textsf{c,}}~Final dimensional radius as a function of contour length in experiments and both types of simulations, for strain rates in the range $\dot{\epsilon}\sim0.3-0.61\,$s$^{-1}$.\ The two dashed lines show the mean radius for experiments (top) and Brownian simulations (bottom), which differ slightly due to the two different methods for estimating~$R$. \textbf{\textsf{d,}} Final dimensional radius $R$ as a function of elastoviscous length $\smash{(B/\mu \dot{\epsilon})^{1/4}}$ in Brownian simulations for varying persistence lengths $\ell_p$, showing a linear dependence in agreement with Eq.~(\ref{eq:scaling}).\ In these simulations, $L=0.6\,\mu$m, and $\dot{\epsilon}\sim10^5$--$10^7\,$s$^{-1}$.\ Also see Supplementary Video 6. \textbf{\textsf{e,}} Collapse of all the data from experiments, {Brownian and non-Brownian }simulations when plotted in dimensionless variables, showing a power-law scaling of the form $R/L\sim \bar{\mu}^{-0.24}$. }\vspace{-0.2cm}
	\label{fig:Radius}
\end{figure*}

We are now in a position to explain the emergence of helicoidal shapes.\ In a strong flow, unstable eigenmodes are planar but can develop and grow in any plane containing the flow axis.\ In addition, dominant modes always come in odd--even pairs with nearly identical growth rates.\ When a straight filament is perturbed, there is thus a strong likelihood for both modes to grow simultaneously. The superposition of two adjacent odd--even planar modes such as those shown in Fig.~\ref{fig:Eigmode}\textbf{\textsf{b}} and growing in different planes produces a coiled three-dimensional conformation that resembles a helix and continues to grow as such, with the two modes interacting as a consequence of geometric nonlinearities.\ A similar mechanism was previously proposed to explain the buckling of elastic rods in soft elastomer matrices \cite{su2014buckling},  though the governing equations and forces at play are very different in that problem.

This modal interaction process can be formalized by deriving a nonlinear amplitude equation of the Ginzburg-Landau form \cite{drazinreid,gugli2012buckling} as we briefly outline (see Supplementary Information for details).\ Close to the onset of buckling, we expand deformations on the basis of the first two linear eigenmodes as they dominate the unstable spectrum: $h_{y,z}(s,t)=A_1^{y,z}(t)\Phi_1(s)+A_2^{y,z}(t)\Phi_2(s)$.\ Retaining nonlinear terms in Eq.~(\ref{eq:lsbt}) and using orthogonality of the linear eigenfunctions with the eigenfunctions of the adjoint linear operator provides a system of coupled nonlinear ordinary differential equations for the unknown time-dependent amplitudes.\ Solutions of this system exhibit amplitude saturation following an initial exponential growth regime, and in sufficiently strong flows always produce 3D helical conformations consistent with observations.\ Remarkably, even planar initial conditions evolve towards helical shapes as nonlinearities force the modes to interact and spontaneously break symmetry.\ This mechanism is illustrated in Fig.~\ref{fig:Eigmode}\textbf{\textsf{b-c}} for a representative elastoviscous number of $\bar{\mu}=1.8\times10^4$  (also see Supplementary Video 9).\ The two dominant linear eigenmodes in this case are plotted in Fig.~\ref{fig:Eigmode}\textbf{\textsf{b}}, where $\Phi^{(1)}$ is found to be odd while $\Phi^{(2)}$ is even.\ A superposition of these modes in orthogonal planes indeed produces a helix in Fig.~\ref{fig:Eigmode}\textbf{\textsf{c}}, showing snapshots from a numerical solution of the weakly nonlinear model in which the spontaneous symmetry breaking is evident. We note that helix formation is also supported by a favorable energetic landscape (see Supplementary Fig.~8): a filament that is forcefully restricted to buckle and compress in two dimensions indeed shows a monotonic growth of its bending energy that is avoided by the provision to coil  \cite{chelakkot2012flow}.

The mechanism by which planar modes interact to create coiled morphologies is corroborated by non-Brownian simulations at the same value of $\bar{\mu}$ in Fig.~\ref{fig:Eigmode}\textbf{\textsf{d}}: in this example, the helical morphology is also seen to emerge from an initially planar buckling mode and has a final shape that resembles the theoretical prediction of Fig.~\ref{fig:Eigmode}\textbf{\textsf{c}}.\ The mechanism is also consistent with the findings of Fig.~\ref{fig:snapshots}, where we observed that helical buckling only occurs at large $\bar{\mu}$: in weak flows, the two dominant eigenvalues are well separated, resulting in the exponential growth of a single dominant mode and emergence of a planar shape. 

\vspace{0.35cm}
\noindent \textbf{\textsf{Radius of helix is independent of filament length}} \\
We now quantify the evolution of the shape during a buckling event from experiments and simulations. Fig.~\ref{fig:Radius}\textbf{\textsf{a}} shows the helix length, estimated as the end-to-end distance $L_{ee}$, and effective radius $R$ as functions of dimensionless time from an experimental realization;\ similar observations are made in simulations (see Supplementary Fig.~4).\ In all cases, $L_{ee}$ decreases and $R$ increases while the helix forms and gets compressed by the flow.\ The nearly linear decrease of $L_{ee}$ with time allows us to extract a characteristic speed $|\dot{L}_{ee}|$ for compression of the helix, which we plot as a function of $\dot{\epsilon}L$ in Fig.~\ref{fig:Radius}\textbf{\textsf{b}}.\ A linear relationship $|\dot{L}_{ee}|\sim 0.7 \dot{\epsilon}L$ is found in both experiments and simulations, with a slope of less than unity that we attribute to the finite elastic resistance of the buckled helical shapes.

In the final stage of compression, the growth of the helix radius slows down as seen in Fig.~\ref{fig:Radius}\textbf{\textsf{a}}, and a nearly steady shape is reached with a roughly constant radius. We measure this final coiling radius from experiments and corresponding simulations (see Methods and Supplementary Information for details) and discuss its dependence on the relevant parameters.\ 
Experiments typically have access to a limited range of strain rates ($\sim 0.4$--$0.6\,$s$^{-1}$), so that variations in $\bar{\mu}$, covering over three decades, are primarily due to variations in filament length $L$.\ Consequently, the radius is first displayed in Fig.~\ref{fig:Radius}\textbf{\textsf{c}} as a function of $L$, keeping all other parameters constant.\ Quite stunningly, a nearly constant value of $R$ is observed, indicating that the final helix radius is largely independent of contour length. The agreement between experiments and simulations for the same conditions is again very good. 

This peculiar result can be rationalized by a scaling theory for the radius of an inextensible helix undergoing compression in flow (see Supplementary Information for details).\ During compression, the pitch of the helix decreases, causing its radius to increase by inextensibility. Balancing the associated viscous dissipation with the rate of change of elastic bending energy during this process yields the simple scaling 
\begin{equation}
R \sim \left(\frac{B}{\mu \dot{\epsilon}}\right)^{1/4}, \label{eq:scaling}
\end{equation}
which is indeed independent of contour length and only weakly dependent on compression rate.\ This result points to the elastoviscous length $(B/\mu \dot{\epsilon})^{1/4}$ \cite{coq2008rotational}, which is the only length scale of the problem besides $L$, as the fundamental scale for the buckling process.\ To test this scaling law and  probe the dependence on flow strength and bending rigidity, we performed additional Brownian simulations in which strain rate and persistence length were varied while keeping $L$ constant.\ The measured radius for three different persistence lengths is displayed in Fig.~\ref{fig:Radius}\textbf{\textsf{d}} as a function of the elastoviscous length and shows a clear collapse of the data corroborating the scaling prediction. Finally, we summarize all the data from experiments and both types of simulations in dimensionless form in Fig.~\ref{fig:Radius}\textbf{\textsf{c}}, where our model predicts\begin{equation}
\frac{R}{L} \sim \bar{\mu}^{-1/4}. \vspace{-0.0cm}
\end{equation}
A similar collapse is found, with some scatter arising from fluctuation-induced defects.\ A numerical fit yields an exponent of $-0.24$ in excellent agreement with the scaling prediction.

\vspace{0.45cm}
\noindent \textbf{\textsf{A generic transition in strain-dominated flows} }\\
 We have elucidated the coiled morphologies of actin filaments in compressional flow through a combination of experiments, simulations, scaling analysis and weakly nonlinear stability theory.\ The two distinct approaches used in numerical simulations highlight the robustness of this phenomenon, in which neither Brownian fluctuations nor a three-dimensional flow field are necessary conditions for helical buckling.\ The stability theory also supports this idea and explained the origin of these structures in a simple two-dimensional stagnation point flow.\ As uncovered in our analysis, the key to helical coiling is the nature of the eigenspectrum associated with the linearized buckling problem, in which dominant eigenmodes come in odd--even pairs with nearly identical growth rates and interact nonlinearly to form helicoidal shapes. Our analysis is an addition to the study of post-buckling mode interactions that are often responsible for non-planar structures \cite{byskov1977mode}. Remarkably, this distribution of eigenvalues is quite generic, and helical buckling has also {recently} been observed for {very flexible} filaments in shear flow (see Supplementary Fig.~10), where the dynamics is more subtle due to the non-stationary base state of a tumbling straight filament. 

The helical buckling instability uncovered here likely serves as an explanation for a number of past experimental observations where helicoidal morphologies were reported but largely overlooked; e.g., during the manufacturing of synthetic wet-spun fibers for cosmetics where long fibers undergo buckling in a compressional flow \cite{mercader2010kinetics}; or in biology, where the sessile protozoan \textit{Vorticella} is known to propel by exploiting the sudden calcium-powered contraction of its slender stalk \cite{ryu2016vorticella}; or during the transport of elastic fibers in turbulent flows \cite{allende2018stretching}. The fundamental mechanism highlighted by our analysis should advance our understanding of these various phenomena and may also be exploited for the controlled microfabrication of chiral objects from one-dimensional elastic filaments.

\vspace{-0.2cm}
\footnotesize

%
%
\section*{Methods}

\noindent \textbf{\textsf{Experiments.\,\,}} The protocol for the assembly of actin filaments is well controlled and reproducible.\ Concentrated G-actin, which is obtained from rabbit muscle cells and purified according to the protocol described in \cite{spudich1971regulation, liu2018morphological}, is placed into F-Buffer (10$\,$mM Tris-Hcl pH$=$7.8, 0.2$\,$mM ATP, 0.2$\,$mM CaCl$_2$, 1$\,$mM DTT, 1$\,$mM MgCl$_2$, 100$\,$mM KCl, 0.2$\,$mM EGTA and 0.145$\,$mM DABCO) at a concentration of 1$\,\mu$M. At the same time, Alexa488 Fluorescent Phalloidin in the same molarity as G-actin is added to stabilize and visualize actin filaments. After 1 hour of  polymerization in the dark at room temperature, concentrated F-actin is diluted 20--50$\times$ for following experiments. 45.5\,\%\,(w/v) sucrose is added to match the refractive index of the PDMS channel ($n=1.41$) in order to get better image contrast.\ The viscosity of the {suspending fluid} with 45.5\,\%\,(w/v) sucrose is $5.6\,$mPa$\cdot$s at $24$$^\circ$C, measured by an Anton Paar MCR 501 rheometer. The filaments obtained by this protocol have a persistence length $\ell_p=17\pm1\,\mu$m and contour lengths in the range $L\sim 20-80\,\mu$m.

A hyperbolic PDMS channel, which has been optimized by taking into account the effect of the {varying} rectangular cross section \cite{zografos2016microfluidic}, is used to provide the background straining flow.\ In this geometry, filaments experience homogeneous strain rates for a certain residence time as they are transported in the center of the channel by a flow focusing mechanism.\ An automated tracking system is used to keep filaments in the visual field of the camera, with an image blur $\lesssim \pm 0.5\,\mu$m due to velocity differences between stage and flow. 
A stable flow is driven by a syringe pump (neMESYS 290N) and particle tracking velocimetry is used to ensure that the velocity profile is in agreement with theoretical predictions \cite{liu2019optimized}.\ 
The lengths of the channel sections with constant strain rate and linearly varying strain rate are $1200\, \mu$m and $800\, \mu$m, respectively, with a channel width of  $789\, \mu$m upstream and $107\, \mu$m at the center.\ The total flow rate $Q$ is in the range of  $3.1-5.5\,$nL/s, which provides average velocities of $\sim100\,\mu$m/s, maximum velocities of  $\sim 1\,$mm/s, {and a range of compression rates $\dot{\epsilon}\sim 0.4-0.6\, s^{-1}$}. 

Images are captured by a CMOS camera (HAMAMATSU ORCA flash 4.0LT, 16 bits) with an exposure time $\Delta t=40\,$ms and are synchronized with the stage displacements through external triggers. The shapes of the actin filaments are extracted through Gaussian blur, threshold, noise reduction and skeletonize procedures in software ImageJ. A customized MATLAB code based on B-spline interpolation is then used to reconstruct the filament centerline along arclength $s$ and to calculate relevant parameters.

\vspace{0.25cm}\noindent \textbf{\textsf{Numerical simulations.\,\,}} We have performed two complementary sets of simulations in two different flow geometries. Full details of the simulation methods are provided in the Supplementary Information. In the Brownian simulations, the filaments are modeled as inextensible Euler--Bernoulli beams, and their dynamics in flow are captured using local slender-body theory, which accounts for drag anisotropy \cite{tornberg2004simulating}. Brownian fluctuations are included and calculated to satisfy the fluctuation--dissipation theorem \cite{manikantan2013subdiffusive,liu2018morphological}. The background flow is chosen to be a purely two-dimensional compressional flow in free space, with the velocity given by $\bu_\infty = (-x,y,0)$. 

In the non-Brownian simulations, the fibers are modeled as a network of Hookean springs that provide structural rigidity to the filaments \cite{nguyen2014hydrodynamics}, with hydrodynamics captured by the method of regularized Stokeslets \cite{Cortez:2005}. The filaments are finitely extensible and approach the limit of inextensibility for very small aspect ratio.\ We performed these simulations in an axisymmetric channel of circular cross-section that provides regions of constant compression and extension as in the experiments \cite{lagrone2019complex}.

\vspace{0.25cm}\noindent \textbf{\textsf{Shape characterization.\,\,}} We track the evolution of the helix shape as a function of Hencky strain $\dot{\epsilon}t$, which is a measure of the accumulated strain experienced by the filaments from the time $t=0$ when their center of mass enters the compressional region. The length is simply estimated by the end-to-end distance $L_{ee}(t)=\|\bx(L,t)-\bx(0,t)\|$ in the plane of motion, where $\mathbf{x}(s,t)$ is a Lagrangian parameterization of the filament centerline with arclength $s\in[0,L]$. Estimating the coil radius is more challenging and is done using two complementary approaches illustrated in Supplementary Fig. 3.\ As experiments only provide shape projections in the $(x,y)$ plane, we estimate the radius in terms of the lateral extent of the filament as $R_\perp(t) =  \left(y_\mathrm{max}(t)-y_\mathrm{min}(t)\right )/2$.\ In simulations, the full filament shape is available and we define an effective radius by fitting the cross-sectional projection in the $(y,z)$ plane with a circle: $R^2_\mathrm{eff}(t) = \langle y(s,t)^2 + z(s,t)^2 \rangle_s$. In both cases, we only consider the central part of the filament where the conformation is mostly helical and omit filament ends. 

In experiments and Brownian simulations, the final helix shape is reached near a Hencky strain of unity, and we estimate the final radius by averaging either $\smash{R_{\perp}(t)}$ or $\smash{R_\mathrm{eff}(t)}$ over $\smash{\dot{\epsilon}t\sim 0.8-1}$. In non-Brownian simulations, filaments typically experience larger Hencky strains, though the key features of the dynamics remain unaltered; in this case, we estimate the final radius by performing the average over $\dot{\epsilon}t\sim 1.6-2$. 

\vspace*{6mm}

\section*{Acknowledgements}
The authors thank Roland Winkler for helping us appreciate the prevalence of helical shapes, as well as Michael Shelley and Eric Shaqfeh for illuminating discussions.\ AL, {BC} and YL acknowledge funding from the ERC Consolidator Grant PaDyFlow (Agreement 682367).\ DS acknowledges funding from a Paris Sciences Chair from ESPCI Paris.\ This work received the support of Institut Pierre-Gilles de Gennes (\'Equipement d'Excellence, ``Investissements d'Avenir", Program ANR-10-EQPX-34).\ JL, RC, LF acknowledge funding from a Grant from the Gulf of Mexico Research Initiative and from National Science Foundation Grant No. DMS-1043626. 

\section*{Author contributions}
BC and DS performed the Brownian simulations, stability analysis and scaling theory. YL, OdR and AL performed experiments. JL, RC and LF performed non-Brownian simulations. All authors contributed to the analysis and interpretation of data and to the preparation of figures. BC, DS, AL, OdR and LF wrote the paper.  %
\section*{Competing interests}
The authors declare no competing interests.

\bibliography{biblio_helical_coiling}

\end{document}


\title{\Large{Supplementary Information} \\ \vspace{0.1cm} \large{Flexible filaments buckle into helicoidal shapes in strong compressional flows\\} \vspace{0.1cm} \normalsize{Chakrabarti et al.} (2019)}
\maketitle

\section{Computational models and methods}
\subsection{Brownian simulations: slender-body theory}

The actin filaments considered in this work have a characteristic diameter $d \sim 8\,$nm and typical lengths in the range of  $L \sim 18-80\, \mu$m. Due to the slenderness of these filaments (aspect ratio $\alpha\equiv d/L\ll 1$), we opt to describe them as space curves parameterized by arc length $s\in[0,L]$, with Lagrangian marker $\mathbf{x}(s,t)$ denoting the position of any material point along the centerline. Noting that the ratio of the energy associated with stretching to that of bending a filament scales as $E_{str}/E_{bend} \sim (L/d)^2 \sim 10^6$, we assume local filament inextensibility, which results in a metric constraint on the Lagrangian marker: $\mathbf{x}_s \cdot \mathbf{x}_s  =1$, where indices denote partial differentiation and $\mathbf{x}_s$ is the local tangent vector. The bending energy of a filament with contour length $L$ is given by:
\begin{equation}
E = \frac{B}{2} \int_{0}^{L}{| {\mathbf{x}_{ss}} | }^2\,\mathrm{d}s,
\label{Eq:bending E}
\end{equation}
where $B$ is the bending rigidity of the filament, which also defines its persistence length $\ell_p=B/k_\mathrm{B} T$. The phalloidin-stabilized actin filaments used in experiments have a persistence length of $\ell_p=17\pm1\,\mu$m \cite{liu2018morphological}.\ We model them as fluctuating Euler-Bernoulli beams whose motion in an imposed flow $\mathbf{u}_\infty(\mathbf{x})$ is described by local slender-body theory for low-Reynolds-number hydrodynamics \cite{johnson1980improved,tornberg2004simulating} as
\begin{equation}
8\pi\mu[\mathbf{x}_t-\mathbf{u}_\infty(\mathbf{x})]=-\Lambda[\mathbf{f}](s), 	\label{Eq:non-localSBT1}
\end{equation}
with
\begin{align}
& \Lambda[\mathbf{f}] = [-c(\mathbb{1} + \mathbf{x}_s\mathbf{x}_s ) + 2(\mathbb{1} - \mathbf{x}_s\mathbf{x}_s )]\cdot \mathbf{f}(s)  \label{Eq:local operator}.
\end{align}
Here, $\mu$ is the suspending fluid viscosity, and $c=-\ln(e \alpha^2)$ is a geometric parameter. $\Lambda$ is the so-called local mobility operator that accounts for drag anisotropy. The force per unit length  $\mathbf{f}(s)$ has contributions from bending and tension forces as well as Brownian fluctuations: 
\begin{equation}
\mathbf{f}(s,t)=B\mathbf{x}_{ssss}-(\sigma(s)\mathbf{x}_s)_s+\mathbf{f}^{br},	\label{Eq:non-local SBT2}
\end{equation}
where $\sigma(s)$ is a Lagrange multiplier that enforces the constraint of inextensibility and can be interpreted as internal tension. The Brownian force density $\mathbf{f}^{br}$ obeys the fluctuation-dissipation theorem \cite{munk2006dynamics}:
\begin{align}
\langle \mathbf{f}^{br}(s,t) \rangle&=\mathbf{0},\\
\langle \mathbf{f}^{br}(s,t)\mathbf{f}^{br}(s',t') \rangle&=2k_\mathrm{B}T\Lambda^{-1}\delta(s-s')\delta(t-t').
\end{align}
Since the filament is freely suspended, we apply force- and moment-free boundary conditions at both filament ends, which translate to: $\mathbf{x}_{sss}=\mathbf{x}_{ss}=\sigma=0$. In contrast to the experiments, the Brownian simulations were preformed in an unbounded domain with a two-dimensional flow. The velocity field used for the simulations mimicks the constant extension and compression rates occurring in the hyperbolic channel of the experiments, and is expressed as
\begin{equation}
\mathbf{u}_\infty(\mathbf{x}) = 
    \begin{dcases}
    \dot \epsilon\, (x,-y,0) & \mbox{for  }0   \leq x \leq l_1, \\
    \dot \epsilon\, (-x,y,0) & \mbox{for  }l_1 \leq x  \leq l_2. \\
    \end{dcases}
\end{equation}
In the above expression $l_1$ and $l_2-l_1$ are the lengths of the extensional and compressional regions, respectively. Conformations from experiments and simulations in the main text were compared at the same value of Hencky strain, or dimensionless time, $\dot{\epsilon} t$.

We non-dimensionalize the governing equations by scaling spatial variables with $L$, time by the characteristic relaxation time $8\pi\mu L^4/B$, the external flow by $L\dot{\epsilon}$, deterministic forces by the bending force scale $B/L^2$  and Brownian fluctuations by $\smash{\sqrt{L/\ell_p}B/L^2}$ \cite{munk2006dynamics}. The dimensionless equations are given as follows:  \vspace{-0.1cm}
\begin{align} 
& \mathbf{x}_t = \bar{\mu }\,\mathbf{u}_\infty(\mathbf{x}(s,t)) - \Lambda[\mathbf{f}](s)	\label{Eq:non-localSBTND1},\\
& \mathbf{f}=\mathbf{x}_{ssss}-(\sigma(s)\mathbf{x}_s)_s+ \sqrt{{L/\ell_p}}\,\boldsymbol{\zeta},	\label{Eq:non-localSBTND2} \\
& \boldsymbol{\zeta} = \sqrt{\frac{2}{\Delta s \Delta t} }\Lambda^{-1/2} \cdot \boldsymbol{w},\vspace{-0.1cm} \label{Eq:non-localSBTND3}
\end{align}
where two dimensionless groups govern the dynamics: the elastoviscous number $\bar{\mu}= 8\pi\mu\dot{\epsilon} L^4/(B c)$ is the ratio of the characteristic flow time scale to the time scale for elastic relaxation of a bending mode, while $L/\ell_p$ compares the filament contour length to its persistence length and measures the magnitude of thermal fluctuations. In Eq.~\eqref{Eq:non-localSBTND3} for the evaluation of the Brownian fluctuations, $\Delta s$ and $\Delta t$ denote the dimensionless spatial and temporal discretizations, respectively, and $\boldsymbol{w}$ is a Gaussian random vector with zero mean and unit variance. Eqs.~(\ref{Eq:non-localSBTND1})--(\ref{Eq:non-localSBTND2}) are numerically integrated in time using an implicit-explicit time-stepping method that treats the stiff linear terms coming from bending elasticity implicitly and non-linear terms explicitly \cite{tornberg2004simulating}.  At every time step, the unknown tensions are obtained by solution of an auxillary tridiagonal linear system that can be derived from the inextensibility condition: $\mathbf{x}_{ts} \cdot \mathbf{x}_s = 0$. Further details of the numerical method can be found in \cite{tornberg2004simulating,liu2018morphological}. All simulations presented here were carried out with $N=128$--$256$ points along the filament arclength.\ Typical time steps for the simulations were of the order of $\Delta t \sim 10^{-12}-10^{-13}$.

\subsection{Non-Brownian simulations} 

\subsubsection{Fiber model as a network of springs}

\begin{figure}[b]
	\centering\vspace{-0.3cm}
	{\label{fig:fiber_side}\includegraphics[height=4.0cm]{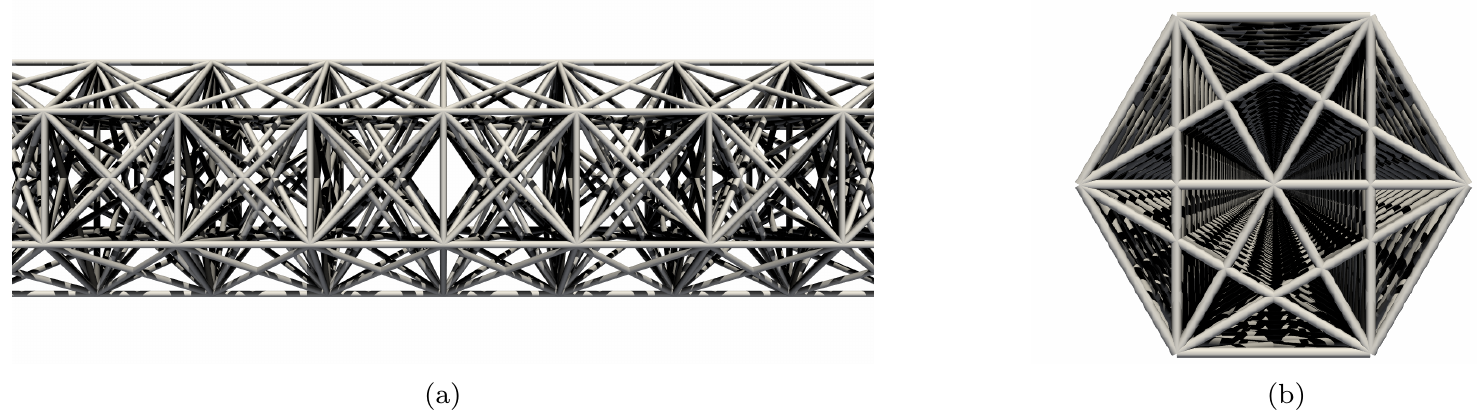}} 
     \vspace{-0.3cm}
	\caption{Schematic of a non-Brownian filament comprised of network of Hookean springs as viewed from the side (a) and from the end (b).}
	\label{fig:fiber}
\end{figure}

The non-Brownian simulations are based on a complementary method for the description of the filaments, which are represented as networks of Hookean springs of stiffness $k$ that provide structural rigidity and bending resistance. The springs connect all pairs of vertices of a hexagonal cross section and additionally link neighboring cross-sections as shown in Figure \ref{fig:fiber}. For an initially straight fiber, the rest length of the spring connecting nodes $i$ and $j$ is set to $\Delta_{ij}$.
The elastic energy resulting from the network of springs is given by:
\begin{align}\label{eq:spen}
E = \frac{1}{2} \sum_{i,j \in \mbox{\scriptsize springs}} k \left(\frac{\ell_{ij}}{\Delta_{ij}} - 1\right)^2 \Delta_{ij},
\end{align}
where $\ell_{ij}$ is the instantaneous length of the spring. We can estimate the bending rigidity $B$, of a filament by bending it to a known curvature $\kappa$ and equating this elastic energy to the bending energy $E_B = B \kappa^2 L/2$, thus providing\begin{align}
B = \frac{2E}{\kappa^2 L}.
\end{align}
We have repeated the above procedure for a number of curvature values to estimate the bending rigidity \cite{peskinlim2004, nguyen2014hydrodynamics} and use this computed value of $B$ to estimate the elastoviscous number $\bar{\mu}$ in the non-Brownian simulations.

\subsubsection{Hydrodynamics: Method of regularized Stokeslets} \label{sec:tube_flow}

%
As discussed previously, the experiments were performed in rectangular hyperbolic channels, while Brownian simulations were carried out in a two-dimensional flow field in absence of any walls. In order to highlight the robustness of helical buckling, we perform non-Brownian simulations in an axisymmetric channel with a circular cross-section. In order to mimic the extensional and compressional flow of the experiment, the dimensionless radius $\rho(z)$ of the channel varies along the axis as follows:
\begin{align} \label{eq:radius}
\rho(z) =\begin{cases}
0.5 \qquad &\mbox{for}\,\,\,\,|z| \geq 2.5, \\
0.286z^3 + 1.63z^2 + 2.78z + 1.74 \qquad & \mbox{for}\,\,\,\,-2.5 < z \leq -1.5, \\
 0.2375\,(z + 2.27)^{-1/2}   \qquad & \mbox{for}\,\,\,\,-1.5 < z \leq 0, \\
 0.2375\,(2.27-z)^{-1/2}     \qquad &\mbox{for}\,\,\,\, 0 < z \leq 1.5, \\
- 0.286z^3 + 1.63z^2 - 2.78z + 1.74 \qquad &\mbox{for}\,\,\,\, 1.5 < z \leq 2.5. \\
\end{cases}
\end{align}
As we demonstrate below, this shape yields in nearly constant extensional rate along the centerline in the central part of the channel.

We represent the fluid velocity induced by the spring forces on the fiber using the method of regularized Stokeslets as discussed in detail in \cite{cortez2001, cortez2005method}. By exploiting the linearity of the Stokes equations, we solve for the tractions on the boundary and compute the  background flow inside the channel using the same method. For an axisymmetric channel, the fluid velocity is expressed linearly in terms of the tractions on the channel wall: 
\begin{align}
\mathbf{u}(\mathbf{x}) = \int_0^L \int_0^{2 \pi} S_\delta \mathbf{F}(\rho(z) \cos(\theta), \rho(z) \sin(\theta), z)\, \rho(z)\, \mathrm{d} \theta\, \mathrm{d}z\,,
\end{align}
where $S_\delta$ is the regularized Stokeslet kernel with regularization parameter $\delta$. $\mathbf{F} = f^r \mathbf{n}_k + f^z \mathbf{e}_3$ are the wall tractions,  where $\mathbf{n}_k=(\cos(\theta),\sin(\theta),0)$. The integral in the azimuthal direction can be evaluated exactly to obtain a formula for the radial and axial velocities:\vspace{-0.05cm}
\begin{align}
\label{eq:ringlet}
\mathbf{u}(\mathbf{x})  = 
\begin{bmatrix}
U^r(\mathbf{x}) \\
U^z(\mathbf{x})
\end{bmatrix}
=
\int_0^L\begin{bmatrix}
 f^r A_r(z) + f^z A_z(z)  \\
f^r B_r(z) + f^z B_z(z) 
\end{bmatrix}\,\mathrm{d}z
=
\int_0^L \mathcal{R}_\delta \mathbf{F} \mathrm{d}z,\vspace{-0.2cm}
\end{align}
where $A_r$, $A_z$, $B_r$, and $B_z$ are comprised of complete elliptic integrals, and we approximate Eq.~\eqref{eq:ringlet} using quadrature (see \cite{ringlets} for more details).
By using this methodology, we are able to discretize the boundary of the tube as a curve of $N_T$ points from
$z=0$ to $z=L$ and $N_{in}$ points discretizing the inlet of the tube from $r=0$ to $\rho(0)$, which yields a significantly reduced total system size of $2(N_T + N_{in}) \times 2(N_T + N_{in})$. 

We then follow the update procedure in \cite{lagrone1}, and compute the flow induced from the no-slip boundary conditions on the tube and the elastic dynamics of the fiber by the following procedure:
\begin{enumerate}
	\item Compute the forces required for the background flow induced by the shape of the tube by solving the linear system\vspace{-0.2cm}
	\[
	\mathbf{u}_{target}(\mathbf{x}_{i}) = \sum_{j=1}^{N_T + N_{in}} \mathcal{R} \mathbf{f}_j h_{tube},\vspace{-0.2cm}
	\]
	where $\mathbf{u}_{target} = \mathbf{0}$ on the tube surface and matches the parabolic flow $\mathbf{u}_{inflow}(r) = U_{max}\left(1 - \frac{r^2}{\rho(0)^2}\right) \mathbf{e}_3$ at the inlet, where $r$ is the radial coordinate with respect to the tube axis and $U_{max}$ is the speed at the center of the tube. We note that this is a dense linear system of size $2(N_T + N_{in}) \times 2(N_T + N_{in})$ that need only be solved a single time. Once we have solved for the forces, Eq.~(\ref{eq:ringlet}) is used evaluate the background flow at any required point.
	
	\item Compute the velocity on all of the fiber nodes resulting from the spring forces using the method of regularized Stokeslets and add the background velocity generated by the boundary forces.
	
	\item Update the position of the fiber by taking a forward Euler time step.
	
\end{enumerate}

\begin{figure}[H]
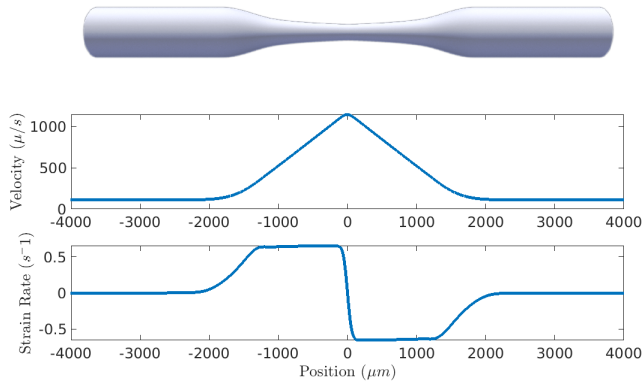

	\centering\vspace{-0.1cm}
	\includegraphics[width=0.48\textwidth]{{{S_SI_FIG2}}} \vspace{-0.1cm}
	\caption{Top: the axisymmetric tube profile. Middle: the simulated axial velocity along the centerline of the tube using the shape defined in Eq.~\eqref{eq:radius}. Bottom: strain rate $\dot{\epsilon}$ along the centerline, which  is the $z$ derivative of the velocity and is approximately constant as desired.}
	\label{fig:flow_linear}
\end{figure}

The described method produces a dimensionless strain rate of $|\dot{\epsilon}| = 4.6$ and the velocity and strain rate profiles along the centerline are shown in Fig.~\ref{fig:flow_linear}.

\subsubsection{Simulation parameters and computed values}

We use non-dimensional parameters in the simulations and convert to physical units using the following characteristic scales for length, time, and force: $L_{scale} = 800\,\mu$m, $T_{scale} = 7.149\,$s, and $F_{scale} = 3.482 \times 10^{-10}\,$N.
These scales where chosen to approximately match experimental values. 
In particular, we normalize the diameter of the experimental channel inlet ($800\, \mu$m) to be $1$,
normalize the experimental inlet velocity ($111.9\, \mu$m/s) to be 1, and normalize the experimental viscosity ($3.89 \times 10^{-3}\,$N\,s/m$^2$) to be 1.

Parameter values used in these non-Brownian simulations are summarized in Table \ref{tab:simulation_parameters}.

\begin{table}[H]
	\centering
	\begin{tabular}{|l|c|c|}
		\hline
		Parameter & Simulation Values & Dimensional Values \\ \hline
		Number of points on tube profile, $N_T$ & 2000 & 2000 \\ 
		Number of points on tube inlet, $N_{in}$ & 400 & 400 \\ 
		Tube grid spacing, $h_{tube}$ & 0.005 & 4 $\mu$m \\ 
		Tube regularization parameter, $\delta_{tube}$  & 0.013 & 10.4 $\mu$m \\
		Inlet speed, $U_{max}$  &	1 &	111.9 $\mu$m/s	\\	
		Shear rate, $\dot{\epsilon}$  & 4.64	& 0.65 s$^{-1}$ \\
		Viscosity, $\mu$ & 	1 &	$3.89 \times 10^{-3}$ N\,s/m$^{2}$ \\
		Filament radius, $r$  &	$1.25 \times 10^{-4}$ & 	0.1 $\mu$m \\	
		Filament length, $L$ & 	0.0134 -- 0.143	&	10.7 -- 114 $\mu$m \\
		Filament discretization size, $h$  &	$1.25 \times 10^{-4}$	&	0.1 $\mu$m \\
		Filament regularization parameter, $\delta$  &	$2.88 \times 10^{-4}$	&	0.23 $\mu$m \\
		Number of filament nodes, $N$ & 738 -- 5550	&	 738 -- 5550	 \\
		Spring constant, $k$ & 0.004 &	$1.40 \times 10^{-12}$ N \\
		Elastoviscous number, $\bar{\mu}$  & $1.22 \times 10^3$ --	$1.08 \times 10^7$  & $1.22 \times 10^3$ --	$1.08 \times 10^7$		\\	
		Bending rigidity, $B$ & 	$2.96 \times 10^{-10}$ -- $2.98 \times 10^{-10}$ &		$6.60 \times 10^{-26}$ -- $6.65  \times 10^{-26}$ N\,m$^2$	\\
		Time step, $\Delta t$  & 	$1 \times 10^{-5}$ & $7.15 \times 10^{-5}$ s \\
		\hline 
	\end{tabular} 
	\caption{Simulation parameters and computed values from non-Brownian simulations.}
	\label{tab:simulation_parameters}
\end{table}

\section{Shape characterization and evolution}

\begin{figure}[H]
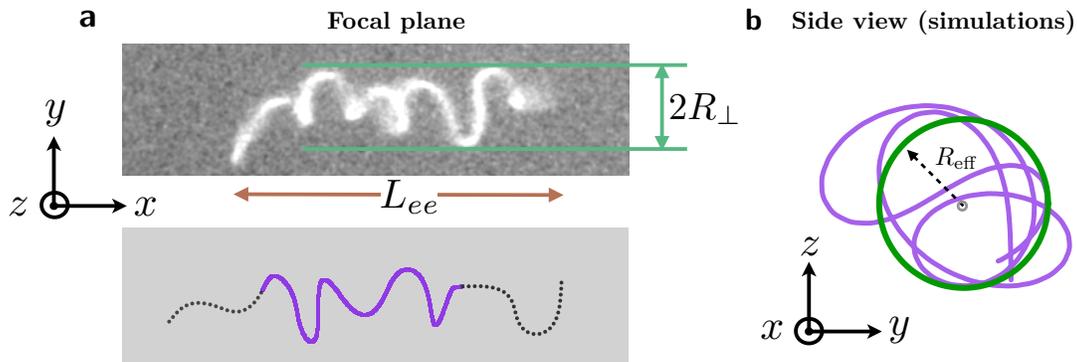

	\centering
	\includegraphics[width=0.8\textwidth]{{{S_rad_measure}}} \vspace{-0.2cm}
	\caption{Characterization of the helical shapes in experiments in simulations. }
	\label{fig:rad_measure}
\end{figure}

Helical shapes in both experiments and simulations are characterized by the effective length and radius of the helices, which are estimated as depicted in Fig.~\ref{fig:rad_measure}. The length is simply estimated by the end-to-end-distance $L_{ee}(t)=\|\bx(L,t)-\bx(0,t)\|$ in the plane of motion, where $\mathbf{x}(s,t)$ is the Lagrangian parameterization of the filament centerline. Estimating the coil radius is more challenging and is done using two complementary approaches.\ As experiments only provide shape projections in the $(x,y)$ plane, we estimate the radius in terms of the lateral extent of the filament as $R_\perp(t) =  \left(y_\mathrm{max}(t)-y_\mathrm{min}(t)\right )/2$.\ In simulations, the full filament shape is available and we define an effective radius by fitting the cross-sectional projection in the $(y,z)$ plane with a circle: $R^2_\mathrm{eff}(t) = \langle y(s,t)^2 + z(s,t)^2 \rangle_s$. In both cases, we only consider the central part of the filament (shown in purple) where the conformation is mostly helical and we omit filament ends. \vspace{-0.2cm}

\begin{figure}[H]
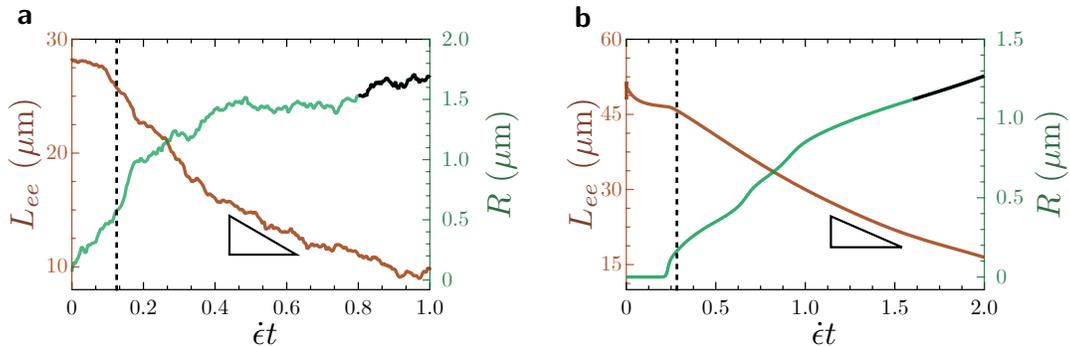

	\centering
	\includegraphics[width=0.8\textwidth]{{{S_end_to_end}}} 
	\caption{Evolution of the end-to-end distance $L_{ee}$ and  coiling radius $R$ as functions of dimensionless time $\dot{\epsilon}t$ in Brownian (\textbf{\textsf{a}}) and non-Brownian (\textbf{\textsf{b}}) simulations. The vertical dashed line shows the onset of buckling.}
	\label{fig:end_to_end}
\end{figure}

Figure~\ref{fig:end_to_end} shows the evolution of the end-to-end distance and helix radius as functions of time in Brownian and non-Brownian simulations. These results should be compared to Fig.~4\textbf{\textsf{a}} of the main text, and show qualitatively similar trends. The nearly linear decrease of $L_{ee}$ with time was used to extract the speed of compression $|\dot{L}_{ee}|$ shown in Fig.~4\textbf{\textsf{b}} of the main draft, and the final helix radius was estimated at the end of compression (black curve). 

\section{Weakly nonlinear theory of mode interactions}
We explain in the main text how the interaction of unstable planar modes can lead to a spontaneous symmetry breaking and formation of helical structures. The eigenspectrum of the linear stability problem, which is quite generic for various problems involving filaments in viscous flows, provides the hint for a representation of deformations during the early stages of buckling.\ In the following discussion, we show how this representation can be used to derive nonlinear amplitude equations for the unstable modes. We start by a brief review of the main results of the linear stability analysis discussed in the main text.

\subsection{Review of the linear stability and unstable eigenspectrum}
\begin{figure}[t]
	\centering
	\includegraphics[width=0.55\textwidth]{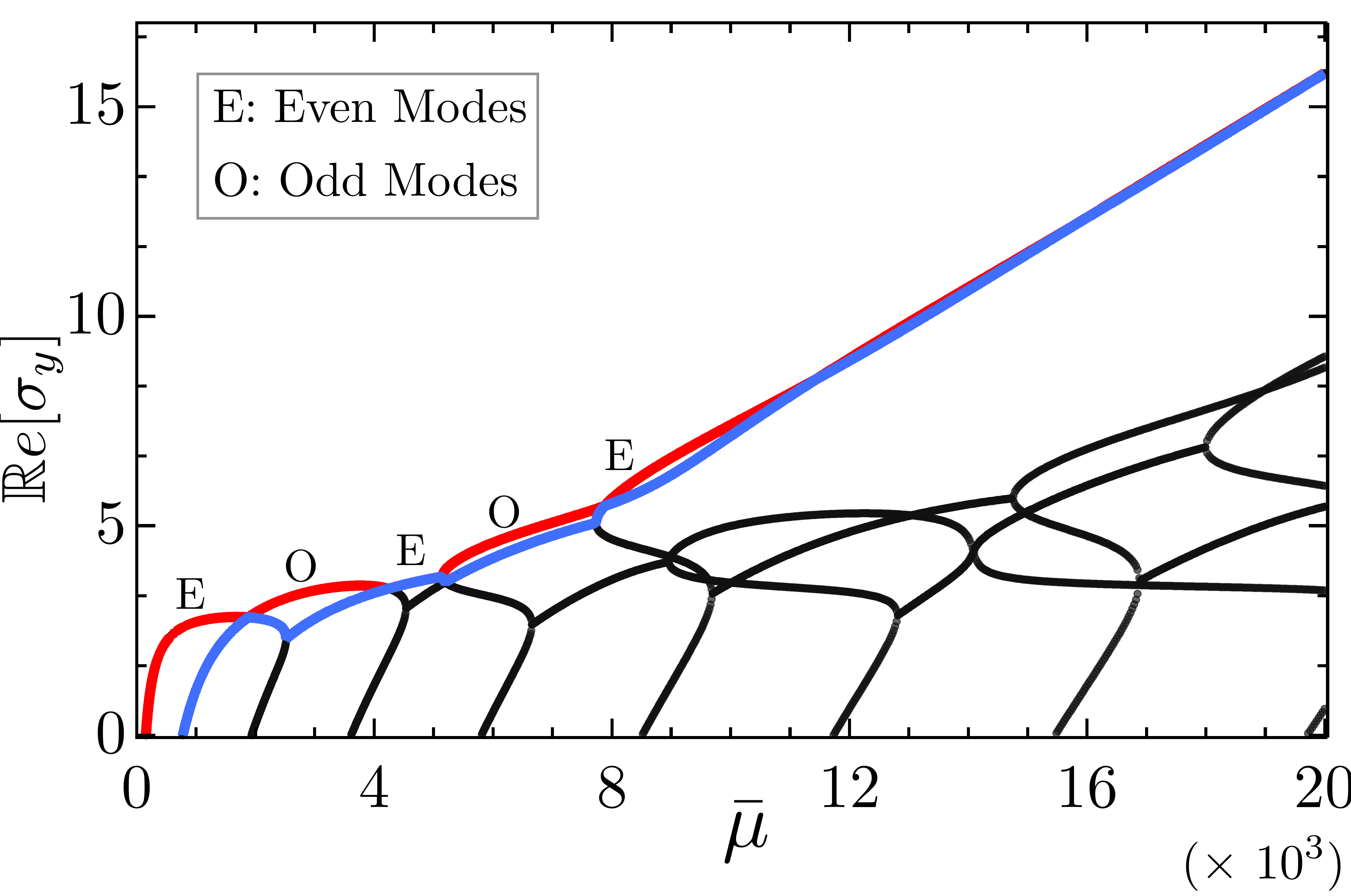}
	\caption{Unstable eigenspectrum of the in-plane problem from the linear stability analysis, where the two dominant modes are highlighted in blue and red. These two modes always come in an odd--even pair.}
	\label{fig:eig}
\end{figure}

Recall that the dynamics of the slender filament in absence of hydrodynamic interactions is governed by the local slender-body equation. In absence of thermal fluctuations and scaling all the time scales with $\dot{\epsilon}^{-1}$, we obtain:
\begin{equation}\label{eq:lsbt}
\bar{\mu}\left(\dot{\mathbf{x}}-\mathbf{u}_{\infty}\right)=\left(\mathbb{1}+\mathbf{x}_{s} \mathbf{x}_{s}\right) \cdot\left[\left(T \mathbf{x}_{s}\right)_{s}-\mathbf{x}_{s s s s}\right],
\end{equation}
where the elastoviscous number $\bar{\mu}$ appears as the sole control parameter. Here, we only retain the logarithmically dominant terms in the equation, under the assumption of a very slender filament. The scalar $T(s)$ is the internal tension that enforces filament inextensibility. Eq.~(\ref{eq:lsbt}) is accompanied by force- and moment-free boundary conditions: $\bx_{sss}=\bx_{ss}=T=0$ at $s=\pm1/2$. In the base state, the compressional flow induces a parabolic tension profile typical of undeformed filaments in linear flows: 
\begin{equation}
T_0(s) = \frac{\bar{\mu}}{4}\left(s^2 - \frac{1}{4}\right).  
\end{equation} 
We perturb the straight configuration as $\mathbf{x}(s,t)=(s,h_y,h_z)$, where $h_y(s,t)$ and $h_z(s,t)$ are in-plane $(x,y)$ and out-of-plane $(x,z)$ shape perturbations, respectively, and are assumed to be small $\mathcal{O}(\varepsilon)$ quantities.\ The  linearized equation of motion for small perturbations is:
\begin{equation}\label{eq:linsb}
\bar{\mu}\,\dot{\bx} = \bar{\mu}\, \bu + T_{0} \bx_{ss} + 2 T_{0,s} \bx_{s} - \bx_{ssss} + \mathcal{O}(\varepsilon^2).
\end{equation}
Resorting to the ansatz of normal modes, we write the perturbations as $\{h_y,h_z\} = \{\Phi_y(s),\Phi_z(s)\}\exp(\sigma t)$, where $\Phi_y$ and $\Phi_z$ are in-plane and out-of-plane mode shapes and $\sigma$ is the complex growth rate. Inserting this form into Eq.~(\ref{eq:linsb}) yields two eigenvalue problems in the $y$ and $z$ directions:
\begin{align}
\bar{\mu}\, (\sigma-1)\, \Phi_y &= \mathcal{L}[\Phi_y], \label{eq:eigx} \\
\bar{\mu}\, \sigma\, \Phi_z &= \mathcal{L}[\Phi_z] \label{eq:eigy} ,
\end{align}
where the linear differential operator $\mathcal{L}$ is defined as
\begin{equation}\label{eq:linop}
\mathcal{L} =  \frac{\bar{\mu}}{4}\left(s^2 - \frac{1}{4}\right)\frac{\partial^2}{\partial s^2} + \bar{\mu} s \frac{\partial}{\partial s} - \frac{\partial^4}{\partial s^4}.
\end{equation}
We observe that the eigenvalue problems in the two orthogonal planes are uncoupled and thus have their own growth rates $(\sigma_y,\sigma_z)$.\ Incidentally, we also see that the two eigenvalue problems are identical under the transformation $\sigma_y'=\sigma_y-1$ which led to the conclusion that at a given value of $\bar{\mu}$, in-plane and out-of-plane mode shapes are identical but have offset growth rates: $\sigma_z=\sigma_y-1$. Fig.~\ref{fig:eig} shows the in-plane unstable eigenvalues as functions of $\bar{\mu}$. 

We emphasize a few important features from this analysis and the associated eigenspectrum that will form the basis of the following nonlinear calculations. In particular, we notice:
\begin{itemize}
	\item At large elastoviscous numbers, the first two eigenvalues $\smash{\{\sigma_y^{(1)},\sigma_y^{(2)}\}}$ dominate the spectrum, so that the corresponding eigenmodes are expected to dictate the emergent morphologies.
	
	\item These two dominant  eigenmodes always come in an odd--even pair, i.e., if $\Phi^{(1)}(s)$ is odd then $\Phi^{(2)}(s)$ is even and vice-versa.
	
	\item The difference in growth rate between these two dominant modes becomes negligible in strong flows.

\end{itemize}

\subsection{Adjoint of the linear operator}
The linear operator $\mathcal{L}$ appearing in the linearized stability problem is not self-adjoint. As a result, the computed eigenfunctions are not orthogonal to each other. In order to proceed with the non-linear theory, we need to find the adjoint $L^\dagger$ of the linear operator and associated adjoint eigenfunctions, which we denote by $\phi^\dagger$. This subtle issue was previously identified by Guglielmini et al.~\cite{gugli2012buckling} in their weakly nonlinear theory for clamped filaments in 2D stagnation point flows. If $v$ is any function satisfying the boundary condition of the linear stability problem with $v_{ss} = v_{sss} = 0$ at $s=\pm1/2$, then the adjoint operator is defined as:
\begin{equation}
\langle u^\dagger,\mathcal{L}v \rangle = \langle L^\dagger u^\dagger, v\rangle,
\end{equation}
where $u^\dagger$ is an adjoint function and the inner-product is simply defined as the integral between $s = \pm 1/2$. Starting from this definition, successive integrations by parts allow us to identify the operator $L^\dagger$ as well as appropriate boundary conditions on $u^\dagger$. The adjoint linear operator in this case is found to be:
\begin{equation}
L^\dagger = \frac{\bar{\mu}}{4}\left(s^2 - \frac{1}{4}\right)\frac{\partial^2}{\partial s^2} - \frac{\bar{\mu}}{2} - \frac{\partial^4}{\partial s^4},
\end{equation}
with boundary conditions:
\begin{align}
u^\dagger_{ss} = u^\dagger_{sss} - \frac{\bar{\mu}}{4}u^\dagger = 0, \ &\text{  at  } \ s= -1/2\\
u^\dagger_{ss} = u^\dagger_{sss} + \frac{\bar{\mu}}{4}u^\dagger = 0, \ &\text{  at  } \ s= 1/2.
\end{align}
The adjoint eigenfunctions then satisfy the eigenvalue problem:
\begin{equation}
L^\dagger \phi^\dagger = \lambda \phi^\dagger,
\end{equation}
with the same boundary conditions.\ By construction, $\mathcal{L}$ and $L^\dagger$ share the same eigenvalues.\ The orthogonality condition can now be written as:
\begin{equation}
\int_{-1/2}^{1/2} \phi^{\dagger(i)} \Phi^{(j)}\,\mathrm{d}s = C \delta_{ij},
\end{equation}
where $C = \langle \phi^{\dagger(i)}, \Phi^{(i)} \rangle$ is a normalization constant.

\subsection{Mode interactions and nonlinear feedback}
We now perform a weakly nonlinear analysis in which we retain the nonlinear term in the slender-body equation. In the early stages of buckling, we make the assumption that the tension profile inside the filament can still be approximated by the tension in the undeformed base state, and thus write the governing equation as
\begin{equation}\label{eq:wnt}
\bar{\mu}\,\dot{\mathbf{x}}=\bar{\mu}\, \mathbf{u}+T_{0} \mathbf{x}_{s s}+2 T_{0, s} \mathbf{x}_{s}-\mathbf{x}_{s s s s}-\bx_s (\bx_s \cdot \bx_{ssss}).
\end{equation}
which can also be written
\begin{equation}\label{eq:nsbt}
\bar{\mu} \,\dot{\bx} = \bar{\mu}\, \bu + \mathcal{L}[\bx] - \bx_s (\bx_s \cdot \bx_{ssss}),
\end{equation}
in term of the linear operator $\mathcal{L}$ defined in Eq.~\eqref{eq:linop}.\ Recall the Lagrangian parametrization of the filament, $\bx = (s,h_y(s,t),h_z(s,t))$.\ Shape perturbations in the transverse direction can be expanded on the basis of the linear eigenfunctions as
\begin{align}
h_y(s,t) &= \sum_{m} A_m^{y}(t) \Phi^{(m)}(s;\bar{\mu}), \\
h_z(s,t) &= \sum_n A_n^{z}(t) \Phi^{(n)}(s;\bar{\mu}).
\end{align}
In the above expansion, $A_m^{y}(t)$ and $A_n^{z}(t)$ are the time dependent amplitude of the perturbations. $\Phi^{(i)}(s;\bar{\mu})$ are the linear eigenfunctions with growth rates $\sigma_i$ at a given $\bar{\mu}$. In the above representation, we have used the fact that the eigenmodes are identical for the in-plane and out-of-plane problems. Since the first two growth rates dominate over the others as seen in Fig.~\ref{fig:eig}, we truncate the infinite sum in the above representation to $m = n = 2$, providing the simple representation
\begin{align}
h_y(s,t) &= A_1^y(t) \Phi^{(1)}(s) + A_2^y(t) \Phi^{(2)}(s), \label{eq:basy}\\
h_z(s,t) &= A_1^z(t) \Phi^{(1)}(s) + A_2^z(t) \Phi^{(2)}(s). \label{eq:basz}
\end{align}
Our aim is to derive ordinary differential equations for the amplitudes of the perturbations that illustrate coupling between unstable modes. To this end, we first explicitly derive the functional form of the nonlinear terms, which we denote by $\boldsymbol{\mathcal{N}}(s,t)=\bx_s (\bx_s \cdot \bx_{ssss})$. We first calculate
\begin{equation}
\beta\equiv \bx_s \cdot \bx_{ssss} = (1,h^y_s,h^z_s) \cdot (0,h^y_{ssss},h^z_{ssss}) = h_s^y h^y_{ssss} + h^z_s h^z_{ssss}.
\end{equation}
Using Eqs.~\eqref{eq:basy}--\eqref{eq:basz} to represent the shapes, we find:
\begin{equation}
\beta = \left(A_1^y \Phi^{(1)}_s + A_2^y \Phi^{(2)}_s\right)\left(A_1^y \Phi^{(1)}_{ssss} + A_2^y \Phi^{(2)}_{ssss}\right) + \left(A_1^z \Phi^{(1)}_s + A_2^z \Phi^{(2)}_s\right)\left(A_1^z \Phi^{(1)}_{ssss} + A_2^z \Phi^{(2)}_{ssss}\right),
\end{equation} 
which can be simplified to:
\begin{equation}\label{eq:beta}
\beta = \left[(A_1^y)^2 + (A_1^z)^2\right] \gamma_1(s) + \left[A_1^y A_2^y + A_1^z A_2^z\right] \gamma_2(s) + \left[(A_2^y)^2 + (A_2^z)^2\right] \gamma_3(s).
\end{equation}
In the above expression, the functions $\{\gamma_1,\gamma_2,\gamma_3\}$ are given by:
\begin{align}\label{eq:coeff}
\gamma_1 &= \Phi_s^{(1)} \Phi_{ssss}^{(2)}, \\
\gamma_2 &= \Phi_s^{(1)} \Phi_{ssss}^{(2)} + \Phi_s^{(2)} \Phi_{ssss}^{(1)}, \\
\gamma_3 &= \Phi_s^{(2)} \Phi_{ssss}^{(2)}.
\end{align}
The transverse components of the nonlinear term $\boldsymbol{\mathcal{N}}(s,t)$ are then obtained as $\mathcal{N}_y = h^y_s \beta$ and $\mathcal{N}_z = h^z_s \beta$, and we emphasize that these nonlinear terms couple in- and out-of-plane modes. 


We now proceed to derive amplitude equations of the Ginzburg-Landau form, first starting with the $y$ component of the slender-body equation. Inserting the expansion for $h_y(s,t)$ into Eq.~\eqref{eq:nsbt} yields
\begin{equation}
\bar{\mu}\Phi^{(1)} \frac{\mathrm{d} A_1^y}{\mathrm{d} t} + \bar{\mu}\Phi^{(2)} \frac{\mathrm{d} A_2^y}{\mathrm{d} t} = \bar{\mu}\left(A_1^y \Phi^{(1)} + A_2^y \Phi^{(2)} \right) + A_1^y \mathcal{L}[\Phi^{(1)}] + A_2^y \mathcal{L}[\Phi^{(2)}] - \mathcal{N}_y.
\end{equation}
Recalling that the eigenfunctions satisfy Eq.~\eqref{eq:eigx}, this simplifies to
\begin{equation}
\bar{\mu}\Phi^{(1)} \frac{\mathrm{d} A_1^y}{\mathrm{d} t} + \bar{\mu}\Phi^{(2)} \frac{\mathrm{d} A_2^y}{\mathrm{d} t} = \bar{\mu} A_1^y \sigma_1 \Phi^{(1)} + \bar{\mu}A_2^y \sigma_2 \Phi^{(2)} - \mathcal{N}_y.
\end{equation}
We now use orthogonality of the above eigenfunctions with their adjoints to obtain amplitude equations. Using the previously defined inner-product, we obtain:
\begin{align}
\frac{\mathrm{d} A_1^y}{\mathrm{d} t} &= \sigma_1 A_1^y  - \frac{1}{\bar{\mu}} \frac{\langle \mathcal{N}_y, \phi^{\dagger(1)}\rangle}{\langle \phi^{\dagger(1)}, \Phi^{(1)} \rangle} , \label{eq:modeeq1} \\
\frac{\mathrm{d} A_2^y}{\mathrm{d}t} &= \sigma_2 A_1^y  - \frac{1}{\bar{\mu}} \frac{\langle \mathcal{N}_y, \phi^{\dagger(2)} \rangle}{\langle \phi^{\dagger(2)}, \Phi^{(2)} \rangle}.
\end{align}
Neglecting nonlinear terms in the above equations recovers the linear stability prediction, where unstable modes grow exponentially with respective growth rates given by the eigenvalues. Nonlinear terms not only introduce additional complexity but also couple the various modes. We can similarly obtain amplitude equations for the out-of plane problem:
\begin{align}
\frac{\mathrm{d} A_1^z}{\mathrm{d} t} &= (\sigma_1-1) A_1^z  - \frac{1}{\bar{\mu}} \frac{\langle \mathcal{N}_z, \phi^{\dagger(1)} \rangle}{\langle \phi^{\dagger(1)}, \Phi^{(1)} \rangle}, \\
\frac{\mathrm{d}  A_2^z}{\mathrm{d} t} &= (\sigma_2-1) A_2^z  - \frac{1}{\bar{\mu}} \frac{\langle \mathcal{N}_z, \phi^{\dagger(2)} \rangle}{\langle \phi^{\dagger(2)}, \Phi^{(2)} \rangle} .  \label{eq:modeeq4}
\end{align}
The fact that the growth rates of the out-of-plane modes are smaller compared to in-plane modes is evident from the the linear terms. All four amplitude equations  \eqref{eq:modeeq1}--\eqref{eq:modeeq4} are coupled to each other through the nonlinear terms and thus need to be integrated numerically.


\subsection{Spontaneous symmetry breaking and helix formation}

\begin{figure}[H]
	\centering
	\begin{overpic}[width=0.83\textwidth]{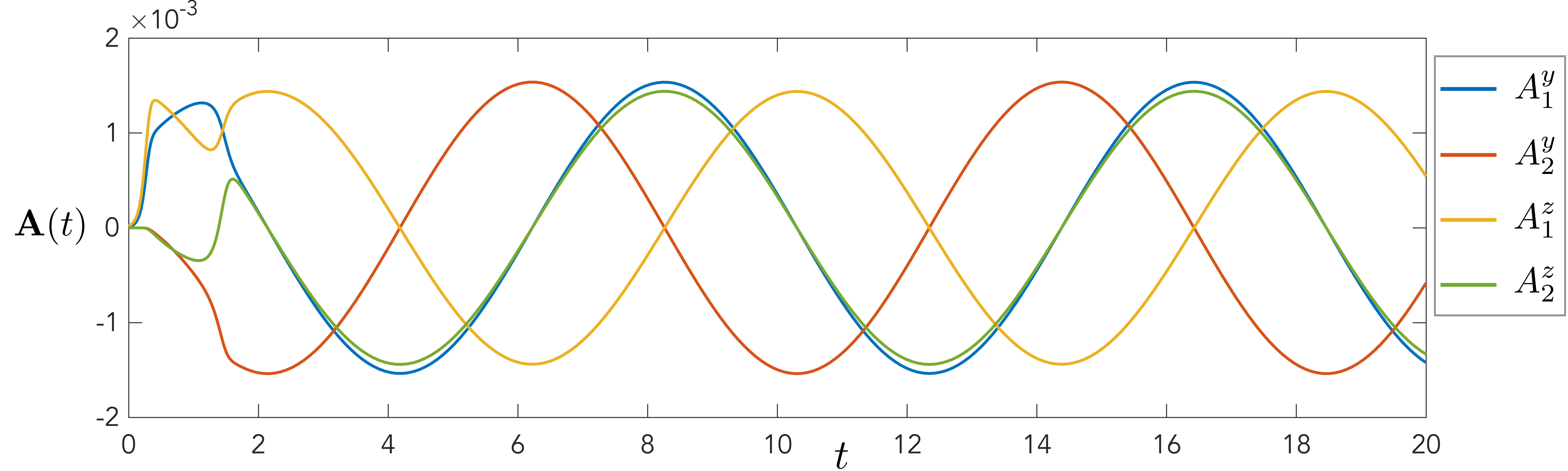}
	\end{overpic}
	\caption{Evolution of the mode amplitudes computed by numerical integration of Eqs.~\eqref{eq:modeeq1}--\eqref{eq:modeeq4} for $\bar{\mu} = 1.8 \times 10^4$ and for the initial condition of Eq.~(\ref{eq:init}). Also see Fig.~3\textbf{\textsf{c}} of the main text  and Supplementary Video 9.}
	\label{fig:amplitudes}
\end{figure}

We recall that the linear stability predicts the same mode shape for the in-plane and out-of-plane problems, and thus predicts  planar conformations for the three-dimensional filament. In order to illustrate the symmetry breaking induced by modal interactions, we thus use a planar conformation as the initial condition for the system of ODEs \eqref{eq:modeeq1}--\eqref{eq:modeeq4}:\vspace{-0.2cm}
\begin{equation}
\mathbf{A}(0) = \epsilon \begin{bmatrix}
1 \\ 0 \\ 1 \\ 0
\end{bmatrix},\label{eq:init}
\end{equation}
where $\epsilon = 10^{-5} \ll 1$. We also choose $\bar{\mu} = 1.8 \times 10^4$ as in the data of Fig.~3 of the main text. A numerical integration of the system of ODEs yields the time-dependent amplitudes shown in Fig.~\ref{fig:amplitudes}. The corresponding shape evolution is illustrated in Fig.~3\textbf{\textsf{c}} of the main text and in Supplementary Video 9.

The time evolution of the amplitudes highlights the subtle role played by the nonlinear feedbacks and modal interactions. We summarize the key results below:
\begin{itemize}
	\item In absence of nonlinearities, the non-zero amplitudes would have amplified exponentially. However the nonlinear terms lead to a saturation of the amplitudes that oscillate over time.
	
	\item Quite remarkably, the amplitudes $A_2^y$ and $A_2^z$ corresponding to the eigenmode $\Phi^{(2)}$ become non-zero and also oscillate in time.
	
	\item A three-dimensional conformation of the filament near the onset of buckling can now be predicted from this weakly nonlinear analysis with $x = s, \  y = A_1^y(t) \Phi^{(1)}(s) + A_2^y(t) \Phi^{(2)}(s)$ and $z = A_1^z(t) \Phi^{(1)}(s) + A_2^z(t) \Phi^{(2)}(s)$. This conformation, as shown in Fig.~3\textbf{\textsf{c}} of the main text and in Supplementary Video 9, clearly results in the formation of a helicoidal shape. 
	
	\item The oscillations in the amplitudes correspond to a slow rotation of the helical conformation, with no change in its overall shape.
	
	\item The time required for saturation of the amplitudes and growth of $A_1^z$ and $A_2^z$ decreases with  $\bar{\mu}$. As a result, experiments and fully nonlinear simulations in strong flows produce helical conformations almost instantaneously after buckling is initiated.
\end{itemize}
The weakly nonlinear analysis highlights how planar modes can interact to give rise to spontaneous symmetry breaking. As mentioned in the main text, the 3D shape predicted from the theory compares well with the simulations at the same elastoviscous number (Fig.~3\textbf{\textsf{c,d}} of the main text). However, the amplitudes predicted by the model do not agree quantitatively with the computed radius of the helical conformation;  we attribute this discrepancy to the approximation made on the internal tension in the theory, which does not preserve filament length.  Predicting the helix radius quantitatively indeed requires enforcement of inextensibility as we explain in the scaling analysis derived in Sec.~V below. Our nonlinear theory does not account for these corrections and can thus only predict emergent shapes in the early stages of buckling. \vspace{-0.2cm}

\subsection{Planar buckling in weak flow}
\begin{figure}[h]
	\centering\vspace{-0.2cm}
	\begin{overpic}[width=0.85\textwidth]{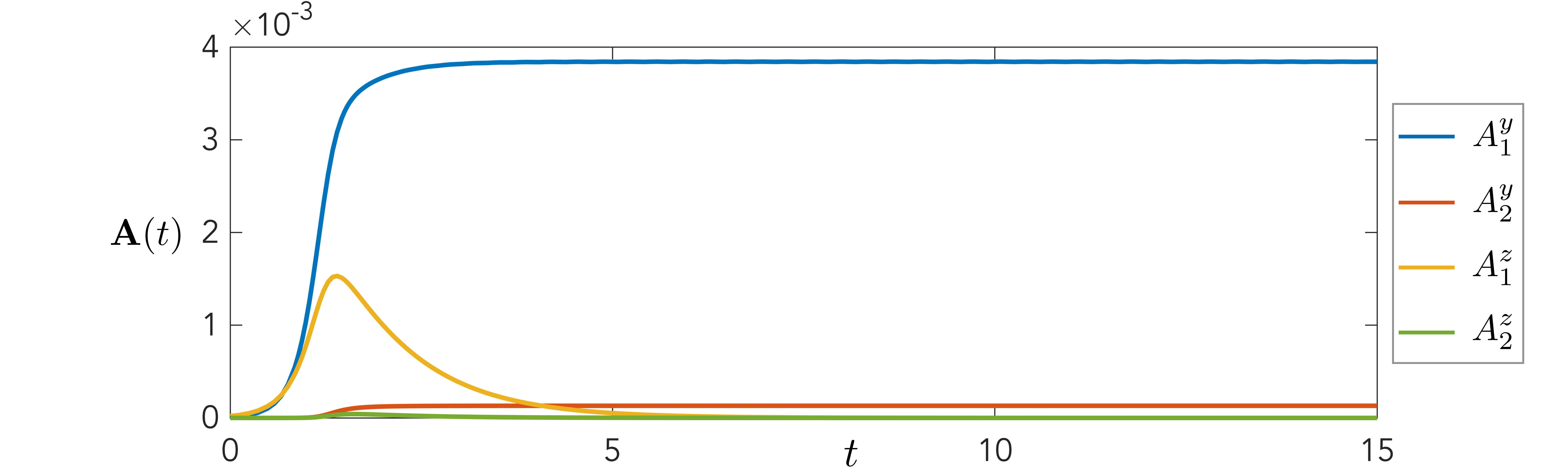}
	\end{overpic}\vspace{-0.2cm}
	\caption{Evolution of the mode amplitudes from a numerical integration of the nonlinear theory at $\bar{\mu} = 7 \times 10^3$, leading to a planar conformation with $A_1^z = A_2^z = 0$ as $t \to \infty$.} \label{fig:amplitudes2}
\end{figure}

We repeat the numerical integration of Eqs.~\eqref{eq:modeeq1}--\eqref{eq:modeeq4} in a weaker imposed flow with $\bar{\mu} = 7 \times 10^3$. The time evolution of the amplitudes in this case is shown in Fig.~\ref{fig:amplitudes2}. As in the previous case, the amplitudes saturate due to nonlinearities. However, we find that  $A_1^z = A_2^z = 0$ as $t \to \infty$ indicating a planar conformation with $z=0$ at long times. This suggests that only planar conformations can arise in weak flows, as the linear growth rates are well separated. This is in accordance with the experiments and simulations discussed in main text. A more systematic exploration of the parameter space using the theoretical model shows that 3D helical shapes can only be sustained for $\bar{\mu} \gtrsim 8 \times 10^3$.

\section{Energy landscape}

\begin{figure}[H]
	\centering
	\includegraphics[scale=0.4]{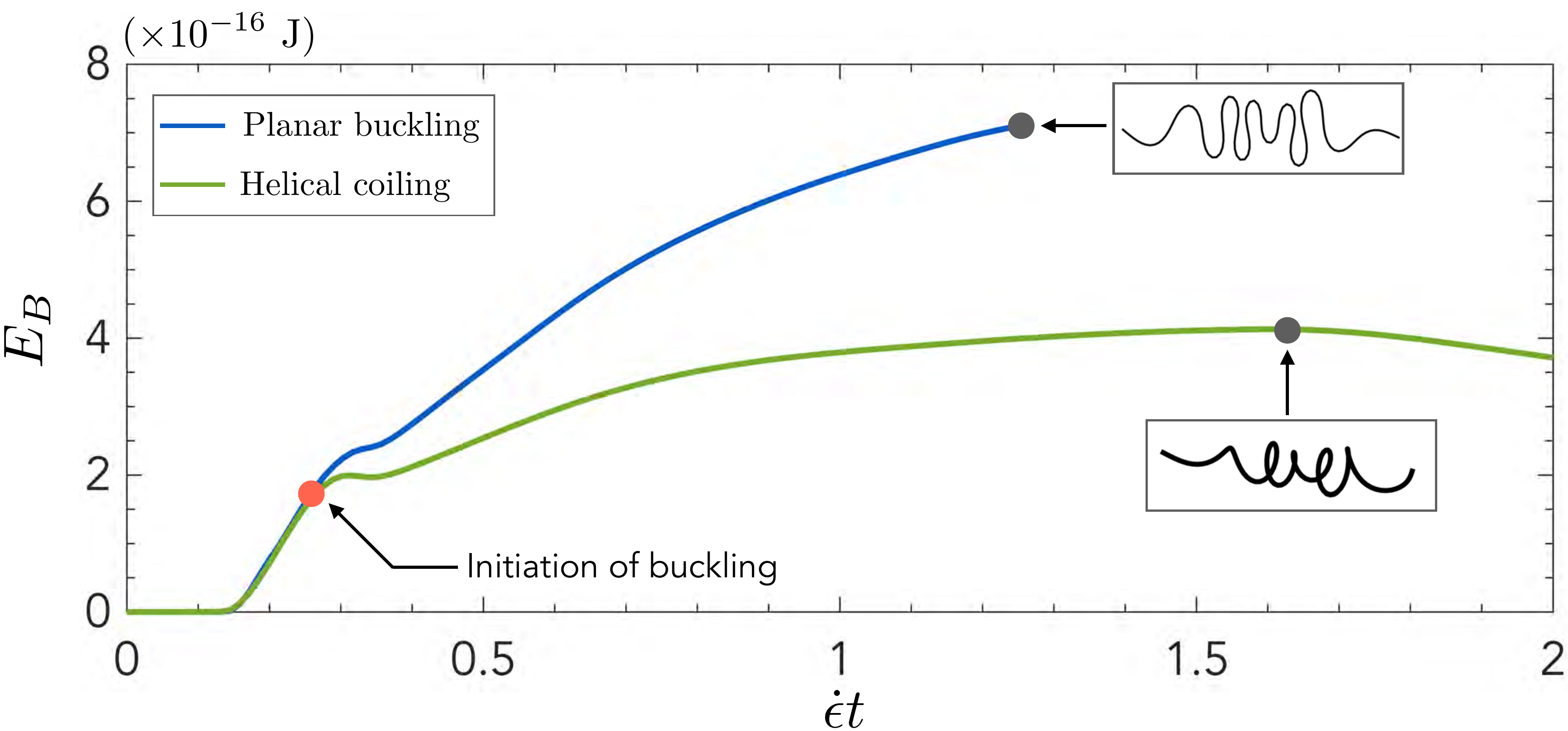}
	\caption{Evolution of the total bending energy in  planar and fully three-dimensional Brownian simulations. The planar simulations were stopped when parts of the filament come in contact. Parameters: $\ell_p/L = 200$ and $\bar{\mu} = 5 \times 10^5$.}
	\label{fig:energy}
\end{figure}

A previous numerical study of helical coiling  \cite{chelakkot2012flow} pointed out that the bending energy of the filament decreases significantly upon initiation of coiling. In order to test this observation, we performed Brownian simulations where we artificially forced the filament to remain planar by restricting the fluctuations or perturbations to two dimensions. After the onset of planar buckling the bending energy of the planar filament grows monotonically as it is compressed by the flow, until contact ultimately occurs. When three-dimensional helical buckling is allowed, the bending energy remains lower and in fact reaches a maximum value as seen in Fig.~\ref{fig:energy} comparing both cases. This suggests that helical buckling is in fact energetically favorable compared to planar buckling in strong flows.  

\section{Scaling analysis for final helix radius}

\begin{figure}[b]
	\centering
	\includegraphics[scale=0.4]{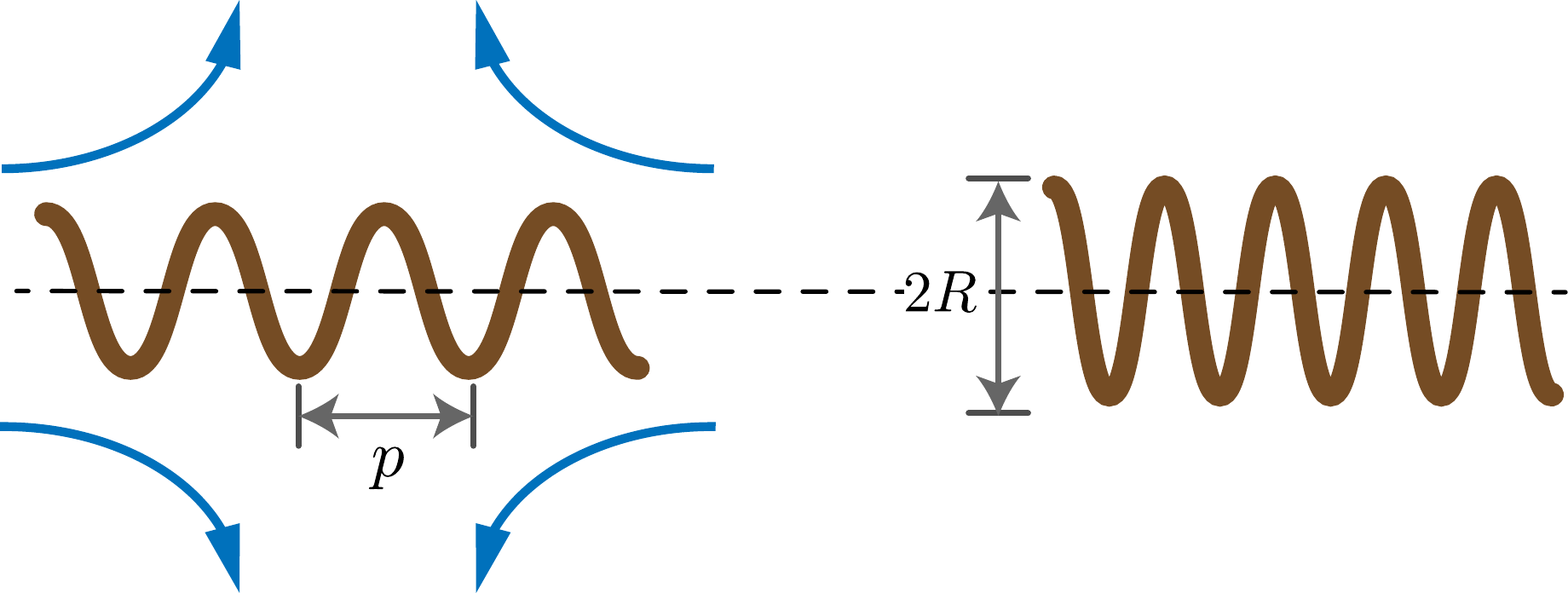}
	\caption{Schematic illustrating the compression of an inextensible ideal helix of radius $R$ and pitch $p$ as used for scaling arguments.}
	\label{fig:compression}
\end{figure}

As seen in both experiments and simulations (see Fig.~4\textbf{\textsf{c}} of the main text), the final helix radius is found to be independent of contour length $L$. We rationalize this peculiar result by developing a scaling theory for the radius of an inextensible helix undergoing compression in flow. For simplicity, we idealize a post-buckling coiled conformation as a perfect helix parameterized by $\theta\in[0,2m\pi]$, where $m$ is the total number of loops. The position of a Lagrangian marker along the curve is then given by $\bx = \left(R \cos \theta, R \sin \theta, \lambda \theta \right)$, where $R$ is the radius and $\lambda$ is related to the pitch $p=2\pi\lambda$. As the helix is compressed by the flow, its pitch decreases while its radius increases as depicted in Fig.~\ref{fig:compression}. This is a consequence of inextensibility, which dictates that\vspace{-0.05cm}
\begin{equation}\label{eq:inex}
R \dot{R} + \lambda \dot{\lambda} = 0.\vspace{-0.05cm}
\end{equation}
During compression, viscous stresses due to the flow balance elastic forces inside the filament. In both experiments and simulations, we observe that $\lambda<R$, which allows us to approximate the dominant viscous force experienced by the helix as $F_v\sim \xi_\perp \dot{p}L$, where $\xi_\perp$ is the coefficient of viscous resistance by unit length in the normal direction and is proportional to the solvent viscosity $\mu$. The associated viscous dissipation can then be estimated as\begin{equation}\label{eq:visdis}
\Phi_v \sim \mu \dot{p}^2 L \sim \mu \dot{\lambda}^2 L, \vspace{-0.05cm}
\end{equation}
and must balance the rate of change of elastic bending energy due to compression. For a helical conformation, the bending energy is given by  $E_B = B L R^2/2(R^2 + \lambda^2)^2$. Using the kinematic relation (\ref{eq:inex}) and the fact that $\lambda<R$, we find its rate of change as\vspace{-0.0cm}
\begin{equation}\label{eq:benen}
\dot{E}_B \sim B L \frac{\lambda \dot{\lambda}}{R^4}. \vspace{-0.0cm}
\end{equation}
In absence of inertia, we seek a balance between $\Phi_v$ and $\dot{E}_B$ \cite{liu2018morphological}. Recalling that $\dot{\lambda}\sim\dot{\epsilon}L$ as found in Fig.~4\textbf{\textsf{b}} of the main text, we arrive at the simple scaling: 
\begin{equation}
R \sim \left(\frac{B}{\mu \dot{\epsilon}}\right)^{1/4} \quad \mathrm{i.e.,}\quad  \frac{R}{L} \sim \bar{\mu}^{-1/4}. 
\end{equation}
This scaling law, which we verify in Fig.~4\textbf{\textsf{d-e}} of the main text, suggests that the helix radius is independent of contour length $L$ and has only a weak dependence on compression rate $\dot{\epsilon}$.

\section{Helical buckling in shear flow}

\begin{figure}[H]
	\centering
	\includegraphics[scale=0.16]{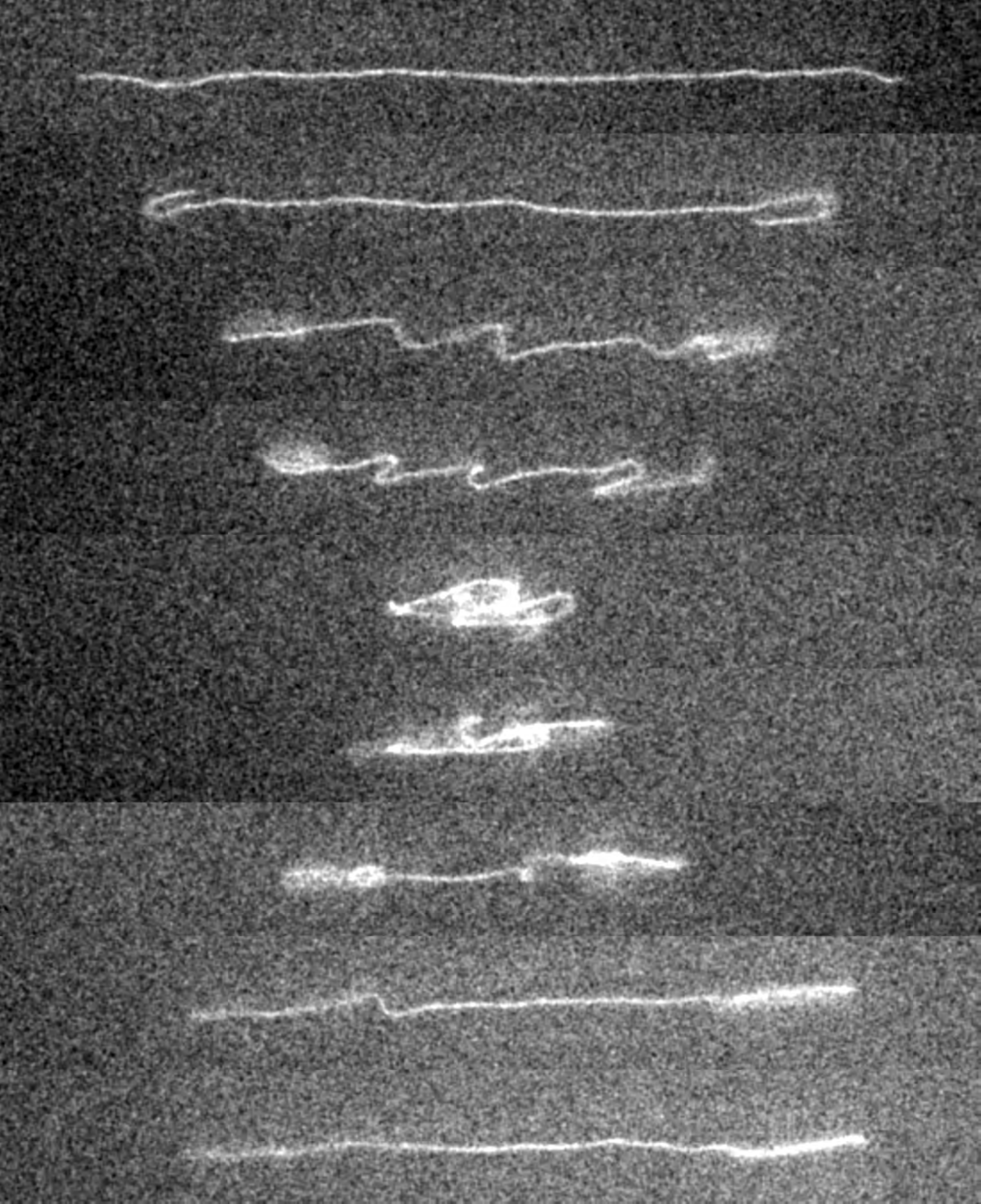}
	\caption{Three-dimensional buckling of a very long actin filament placed in a linear shear flow.}
	\label{fig:shearflow}
\end{figure}

To illustrate the generality of the helical buckling mechanism, we performed additional experiments using the same actin filaments in shear flow, using the experimental setup of Liu et al.~\cite{liu2018morphological}. Large values of the elastoviscous number can be achieved using long filaments, and three-dimensional shapes are indeed observed in this regime as shown in Fig.~\ref{fig:shearflow}. The helicoidal shapes in this case are not as regular as in purely compressional flow, and are also unsteady as the filaments undergo tumbling by a tank-treading motion. 

\section{List of supplementary videos}

\begin{itemize}
	\item  \textbf{Video 1:} Experiment with $\ell_p/L = 1.47$ and $\bar{\mu}=1.2 \times 10^3$, corresponding to the first row of Fig.~1\textbf{\textsf{a}} of the main text. 
	
	\item  \textbf{Video 2:} Experiment with $\ell_p/L = 0.634$ and $\bar{\mu}=2.9 \times 10^4$, corresponding to the second row of Fig.~1\textbf{\textsf{a}} of the main text.

	\item \textbf{Video 3:} Experiment with $\ell_p/L = 0.311$ and $\bar{\mu}=8.8 \times 10^5$, corresponding to the third row of Fig.~1\textbf{\textsf{a}} of the main text.

	\item \textbf{Video 4:} Experiment with $\ell_p/L = 0.254$ and $\bar{\mu}=1.8 \times 10^6$, corresponding to the fourth row of Fig.~1\textbf{\textsf{a}} of the main text. 
	
	\item \textbf{Video 5:} Brownian simulation with $\ell_p/L = 0.254$ and $\bar{\mu}=1.8 \times 10^6$, corresponding to the fourth row of Fig.~1\textbf{\textsf{a}} of the main text.

	\item \textbf{Video 6:} Brownian simulation with $\ell_p/L = 30$ and $\bar{\mu} = 2 \times 10^5$.
	
	\item \textbf{Video 7:} Non-Brownian simulation with $L = 0.026$ and $\bar{\mu} = 1.8 \times 10^4$.
		
	\item  \textbf{Video 8:} Non-Brownian simulation with $L = 0.038$ and $\bar{\mu} = 6.5 \times 10^4$.
	
	\item \textbf{Video 9:} Emergence of a 3D helical shape from a planar linear eigenmode as predicted by the weakly nonlinear theory at $\bar{\mu} = 1.8 \times 10^4$.
\end{itemize}

\bibliography{supplement}